\documentclass[tightenlines,superscriptaddress,twocolumn,preprintnumbers,amssymb,nofootinbib]{revtex4-1}

\usepackage{amsmath, amsthm, amssymb} 
\usepackage{amssymb, physics, bm}
\usepackage{subfigure}
\usepackage{epsfig}
\usepackage{graphicx}
\usepackage{wrapfig}
\usepackage{color}
\usepackage{tikz}
\usepackage[normalem]{ulem}
\usetikzlibrary{arrows,shapes}
\usetikzlibrary{trees}
\usetikzlibrary{matrix,arrows} 			
\usetikzlibrary{positioning}				
\usetikzlibrary{calc,through}				
\usepackage{pgffor}							

\newcommand{\xx}{\mathbf{x}}
\newcommand{\yy}{\mathbf{y}}
\newcommand{\pp}{\mathbf{p}}

\newcommand{\dmu}{\partial_{\mu}}

\newcommand{\lag}{\mathcal{L}}
\newcommand{\ham}{\mathcal{H}}

\DeclareMathOperator{\diag}{diag}
\DeclareMathOperator{\timeorder}{T}
\newcommand{\MS}{\overline{\mathrm{MS}}}

\newcommand{\chiM}{$\chi M$}

\newcommand{\expect}[1]{\left\langle #1 \right\rangle}

\newcommand{\ddpp}{\int \frac{ \dd^3 p }{ (2\pi)^3 }}

\usepackage{siunitx}

\newcommand{\beq}{\begin{equation}}
\newcommand{\eeq}{\end{equation}}
\newcommand{\bqa}{\begin{eqnarray}}
\newcommand{\eqa}{\end{eqnarray}}

\newcommand{\os}{\text{\tiny OS}}
\newcommand{\ms}{\overline{\text{\tiny MS}}}

\def\square{\vcenter{\vbox{\hrule height.4pt
          \hbox{\vrule width.4pt height4pt
          \kern4pt\vrule width.3pt}\hrule height.4pt}}}

\begin{document}

\title{Thermodynamics and phase diagrams of Polyakov-loop extended chiral models}

\author{{\AA}smund Folkestad}
\email{afolkest@mit.edu}
\affiliation{Department of Physics, Faculty of Natural Sciences, NTNU, 
Norwegian University of Science and Technology, H{\o}gskoleringen 5,
N-7491 Trondheim, Norway}
\affiliation{Center for Theoretical Physics,
Massachusetts Institute of Technology, 
Cambridge, MA 02139 USA}

\author{Jens O. Andersen}
\email{andersen@tf.phys.ntnu.no}
\affiliation{Department of Physics, Faculty of Natural Sciences, NTNU, 
Norwegian University of Science and Technology, H{\o}gskoleringen 5,
N-7491 Trondheim, Norway}

\date{\today}

\begin{abstract}
We study the thermodynamics and phase diagrams
of two-flavor quantum chromodynamics using the Polyakov-loop
extended quark-meson (PQM) model and the 
Pisarski-Skokov chiral matrix ($\chi M$) model~\cite{PisarskiSkokov}. 
At temperatures up to $T\approx2T_c$ and baryon chemical potentials up to $\mu_B=400\ \rm{MeV}$, both models show reasonable agreement 
with the pressure, energy density, and interaction measure as calculated on the lattice. 
The Polyakov loop is found to rise significantly 
faster with temperature in models than on the lattice. 
In the low-temperature and high baryon density regime, the two models predict different states of matter; The PQM model predicts a confined and chirally restored phase, while the $\chi M$ model predicts a deconfined and chirally restored phase. At finite isospin density and zero baryon density, the onset of pion condensation at $T=0$ is at
$\mu_I={1\over2}m_{\pi}$, and the transition is second order at all temperatures.
The transition temperature for pion condensation coincides with that of 
the chiral transition for values of the isospin chemical potential
larger than approximately $110\ \rm{MeV}$. In the $\chi M$ model they also coincide with the transition temperature for deconfinement.
The results are in good overall agreement with recent lattice simulations of the $\mu_I$--$T$ phase diagram.
\end{abstract}


\maketitle

\section{Introduction}
The first phase diagram of QCD appeared in the 1970s, and at the time
it was thought that it consists of two phases: A hadronic low-temperature
phase and a high-temperature phase of deconfined quarks and gluons.
Today, the conjectured phase diagram in the $\mu_B$--$T$ plane is
far more complicated. In particular, it is believed that the deconfined
quark phase at high density and low temperature consists of
various color-superconducting phases, with different patterns
of spontaneous symmetry breaking. Some of these phase may even be inhomogeneous, see Refs.~\cite{wil,alford,hatsuda}
for reviews. However, only a few exact results
are known: due to asymptotic freedom we know that at asymptotically high temperature QCD is in
a plasma phase of weakly interacting quark and 
gluons. Similarly, due to asymptotic freedom and the existence of an attractive interaction via 
one-gluon exchange, we have a superconducting color-flavor locked phase at asymptotically high densities. A severe problem in the efforts to map out the QCD phase diagram between these asymptotic regions is that one cannot use lattice simulations at finite baryon
density due to the so-called sign problem. The fermion determinant is complex and one cannot
apply standard Monte-Carlo techniques based on importance sampling.
Thus one typically has to resort
to low-energy effective models such as the quark-meson (QM) model, the Nambu-Jona-Lasinio (NJL) model, and their
Polyakov-loop extended versions~\cite{fuku00}.

There are other external parameters that one can introduce in addition to the temperature
and the baryon chemical potential.
For example, one can add a (strong) magnetic background field $B$. QCD 
in strong magnetic fields  is relevant in e.g. heavy-ion collisions~\cite{mag1,mag2,mag3} and
compact stars~\cite{duncan}. One can also use an independent chemical potential
$\mu_f$ for each quark flavor. In two-flavor QCD, this implies
that one uses $\mu_u$ and $\mu_d$, or equivalently, $\mu_B$ in 
addition to the isospin chemical potential $\mu_I$.
Isospin asymmetry and the possibility of Bose condensation of charged pions may also be
relevant to compact stars.
An advantage of QCD in a magnetic field or at finite isospin
density (but at zero $\mu_B$) is that there is no sign problem, 
and one can therefore use standard Monte-Carlo techniques to
study the phase diagram of these systems.
This opens up for the possibility to confront results from 
model calculations with those of the first-principle method of lattice QCD.

In this paper, we study the thermodynamics of two-flavor QCD using the Polyakov-loop
extended quark-meson (PQM) model and the Pisarski-Skokov chiral matrix model ($\chi M$) 
\cite{PisarskiSkokov} adapted for two flavors
in the mean-field approximation. In this approximation, the mesonic fields are treated at tree level
while the fermion fields are integrated over in the Gaussian approximation. 
At one loop, dropping the mesonic fluctuations is equivalent to working in the large-$N_c$ limit. Sometimes
the no-sea approximation is made, which simply means that one discards the fermionic quantum
fluctuations. However, one should keep vacuum fluctuations since there is no apriori reason to omit them, not even in low-energy models of an underlying theory. Secondly,  it turns out that the
inclusion of quantum fluctuations change the order of a phase transition in some cases; in 
the two-flavor QM model the chiral transition changes from first order to second in the chiral limit, 
showing the importance of keeping them.
Moreover, in almost all mean-field calculations to date, the parameters
of the Lagrangian are determined at tree level. This is inconsistent since the
effective potential has been determined in the one-loop large-$N_c$ approximation.
The parameters should always be determined at the same level of accuracy as the effective
potential, otherwise erroneous results may occur. For example, the onset of pion condensation at $T=0$
takes place when the isospin chemical potential equals half the pion mass. This exact result is
only reproduced if the matching of the parameters is done in a consistent manner.

The paper is organized as follows. In Sec.~\ref{sec:cent_sym}, we discuss various aspects
of the Polyakov loop and its properties. In Sec.~\ref{sec:gluon}, the gluonic
sector of the PQM and $\chi M$ models is reviewed, while in 
Sec.~\ref{sec:chiral} their chiral sectors are discussed. 
Complications related to minimizing the effective potential at nonzero baryon chemical potential is discussed in Sec.~\ref{sec:minProcedure}. In Sec.~\ref{sec:thermo}, we
present the main result of the paper, namely the thermodynamic functions
and the phase diagram in the $\mu$--$T$ and $\mu_I$--$T$ planes. Our results are compared with lattice results.
In Sec. VII, we summarize and conclude. Four appendices are devoted
to technical details.

\section{Center symmetry and the Polyakov loop}
\label{sec:cent_sym}
Let $T^a$ be the generators of $SU(N_c)$ in the fundamental representation.
A  gauge transformation of the QCD gluon field $A_{\mu}=A_{\mu}^aT^a$
is of the form
\begin{equation}
  A_{\mu}(x) \rightarrow \Omega(x)A_{\mu}(x) \Omega^{\dag}(x) -
  \frac{ i }{ g } \left[\dmu \Omega(x)\right]\Omega^{\dag}(x)\;,
\label{eq:gaugeReminder}
\end{equation}
for any $\Omega(x)$ in the fundamental representation of $SU(N_c)$. This transformation leaves the gluonic Lagrangian 
invariant, and is thus a symmetry of the action of the pure gauge theory. 
However, when studying QCD at finite temperature $T=\beta^{-1}$ in the imaginary time formalism, 
choosing a generic $\Omega(\xx, \tau)$ ruins the periodicity of $A_\mu(\xx, \tau)$
in imaginary time $\tau$,
as required for the field configurations summed over in
the partition function ${\cal Z}$.\footnote{When working with imaginary time $t=-i\tau$, we redefine fields as $A_\mu(\xx, -i\tau) \rightarrow A_\mu(\xx, \tau)$.}
Restricting ourselves to transformations that satisfy
\begin{equation}
	\Omega(\xx, \tau) = \Omega(\xx, \tau + \beta)\ , \label{eq:periodicGauge}
\end{equation}
avoids the problem, but there is a larger group of symmetries
that preserves the imaginary time periodicity of $A_\mu$. Consider instead a generic gauge transformation
that satisfies 
\begin{equation}
  \Omega(\xx, \tau+\beta) = G(\xx, \tau) \Omega(\xx, \tau)\ ,
  \label{eq:twistedGauge}
\end{equation}
for some $G(\xx, \tau) \in SU(N_c)$. Let $A'_{\mu}$ be the transformed
field.
We then get
\bqa\nonumber
A^{\prime}_{\mu}(\xx, \tau+\beta) &=&
G(\xx, \tau)A'_{\mu}(\xx, \tau)
  G^{\dag}(\xx, \tau)
\\ &&
  - \frac{ i }{ g }\left[\dmu G(\xx, \tau)\right]
  G^{\dag}(\xx, \tau)  \;.
  \eqa
If $G(\xx, \tau)$ is constant in space and imaginary time and
commutes with $A^{\prime}_\mu$ for all $(\xx, \tau)$, then the
gauge field is periodic.
Since $\Omega(\xx, \tau)$ is a matrix in the fundamental representation of
$SU(N_c)$, which is irreducible, $G$ is proportional to the identity matrix by
Schur's lemma. Let $G = \lambda I_{N_c}$, where $I_{N_c}$ is the $N_c \times N_c$
identity matrix and $\lambda \in \mathbb{C}$. Since we know that $G\in SU(N_c)$,
we have that $\lambda = \lambda_n$ is one of the $N_c$-th roots of unity, and all
possible matrices $G$ are given by 
\begin{equation}
G_n = \lambda_n I_{N_c} = e^{-2\pi i n/N_c}I_{N_c}\ , \quad n=0, \ldots, N_c - 1\ . 
\end{equation}
Clearly $\left\{G_n\right\}$ forms a finite group that is isomorphic to
$\mathbb{Z}_{N_c}$, and it is the center group of $SU(N_c)$.
We refer to aperiodic gauge transformations, characterized by $G_n\neq I_{N_c}$, as twisted gauge transformations or center transformations. 

In pure gauge theory, the expectation of the Polyakov loop
operator is  an  order  parameter  for
deconfinement~\cite{SvetitskyYaffe1,SvetitskyYaffe2, McLerranSvetitsky1}. For QCD
with dynamical quarks, it is an approximate order parameter, 
similar to the chiral condensate.
The thermal Wilson line is given by
\begin{equation}
L(\xx) = \timeorder_{\tau} \exp\left[ig \int_0^{\beta} \dd \tau A_4^a(\xx, \tau) 
T^a\right]\;,
\label{eq:PolyakovLoop}
\end{equation}
where $\timeorder_{\tau}$ denotes time ordering. Here $A_4^a$ is the Euclidean temporal gauge field that replaces the Minkowski temporal gauge field in the Euclidean QCD action through the replacement $A_0 \rightarrow iA_4$, with $A_4$ Hermitian \cite{skokovrev}. 

The Wilson line is not
invariant under (periodic) gauge transformations $\Omega(\xx,\tau)$, but
transforms as $L(\xx)\rightarrow\Omega(\xx,\beta)L(\xx)\Omega^{\dagger}(\xx,0)$.
Taking the trace over color indices, however, yields a gauge-invariant operator, 
which is the definition of the 
Polyakov loop operator 
\begin{equation}
\Phi(\xx) = \frac{ 1 }{ N_c } \tr_c L(\xx)\;.
\end{equation}
Under twisted gauge transformations, the 
Polyakov loop operator transforms nontrivially,
$\Phi(\xx)\rightarrow \Phi^{\prime}(\xx)=\frac{ 1 }{ N_c }\Tr\left[\lambda_n\Omega(0, \xx) L'(\xx) \Omega^\dag(0, \xx)\right] = \lambda_n \Phi(\xx)$.
Thus, we see that the Polyakov loop is gauge invariant ($n=0$), but
not center symmetric. Therefore
the thermal expectation value of the Polyakov loop operator 
transforms as 
\bqa
\langle\Phi\rangle
\rightarrow \lambda_n\langle\Phi\rangle\;.
\eqa
Thus if $\langle\Phi\rangle\neq0$, the center symmetry is spontaneously broken. 

While $\Phi$ is related to the free energy of a heavy quark, 
the conjugate Polyakov loop $\bar{\Phi}$ is the analogue of $\Phi$ for antiquarks, 
and it is obtained from $\Phi$ by replacing $T^a \rightarrow -(T^a)^* = -(T^a)^T$, 
ie.\ by switching to the complex conjugate representation. This is shown in Appendix~\ref{heavyprop}.
After using that $\tr X = \tr X^T$ for any matrix $X$, we find 
\bqa\nonumber
\bar{\Phi}&=&{1\over N_c}{\rm tr}_c 
\left[\timeorder_{\tau} e^{ig \int_0^{\beta} \dd \tau A_4^a(\xx, \tau) T^a}
\right]^{\dagger}
\\&=&
{1\over N_c}{\rm tr}_cL^{\dagger}(\xx) = \Phi^\dag\;,
\label{reala}
\eqa
if the fields $A_4^a(\xx, \tau)$ are real. More generally, if $A_4^a(\xx, \tau)$ are complex, we find 
\bqa\nonumber
\bar{\Phi}&=&{1\over N_c}{\rm tr}
\bar{\timeorder}_{\tau} e^{-ig \int_0^{\beta} \dd \tau A_4^a(\xx, \tau)}
\\&=&
{1\over N_c}{\rm tr}_c\bar{L}(\xx)\;,
\eqa
where $\bar{\timeorder}_{\tau}$ denotes anti-time ordering. 
In full QCD one would not have to worry about complex $A_4^a$ fields, since any particular $A_4^a$ configuration occurring in the path integral should be real. Thus, when averaging over all field configurations in full QCD one has $\expect{\Phi^\dag} = \expect{\bar{\Phi}}$.
However, the $A_4^a$ background fields we will deal with in the effective models should be thought of as mean fields $\expect{A_4^a}$, which in QCD with $\mu_B\neq0$ are obtained by averaging real field configurations with potentially complex weights $e^{-S_E}$. 
The distinction between $\Phi^\dag$ and $\bar{\Phi}$ matters only when dealing with $\Phi(\expect{A_4})$ and $\bar{\Phi}(\expect{A_4})$, i.e.\ the loops evaluated at the mean field $\expect{A_4}$, rather than the expectation values of the loops themselves. In summary: $\bar{\Phi}(\expect{A_4}) \neq \Phi^\dag(\expect{A_4})$ even though $\expect{\bar{\Phi}} = \expect{\Phi^\dag}$. In the effective models, the former occurs. 
We will return to this in the following when we discuss minimization of the effective potential at $\mu_B\neq0$.
Using the expression for $\bar{\Phi}$ is anyway always correct.
If we use $\Phi^\dag$ in place of $\bar{\Phi}$ without an expectation value, it is implied that $A_4^a$ is real.

In the original paper by McLerran and Svetitsky \cite{McLerranSvetitsky1}, it was argued that 
\begin{multline}
    e^{-\beta F(\xx_1, \ldots, \xx_n, \yy_1, \ldots, \yy_{\tilde{n}})} =\\ \expect{\Phi(\xx_1)\ldots \Phi(\xx_n) \Phi^\dag(\yy_1)\ldots \Phi^\dag(\yy_{\tilde{n}})}\ ,
\end{multline}
where $F$ is the color-averaged free energy of a configuration of quarks located at $\xx_1, \ldots \xx_n$ and antiquarks located at $\yy_1, \ldots \yy_{\tilde{n}}$.
We can thus interpret $-T \ln \langle\Phi({\bf 0})\rangle$ as the free energy
of a single quark and $-T \ln \langle\Phi^{\dagger}(\xx)\rangle$ as the free energy
of a single antiquark. If $\langle\Phi\rangle=0$, this implies that the free energy
of a quark is infinite, or that quarks are confined.

Another way  to think of confinement is in terms of the quark
propagator. The
Polyakov loop is proportional to the expectation value of
the traced
propagator of a heavy quark analytically continued to imaginary time.
In Appendix \ref{heavyprop}, we show that
\bqa\nonumber
\bra{q_a(\xx,0)}\ket{q_a(\xx,-i\beta)} 
&=& \left[G(\xx, -i\beta; \xx, 0)\right]_{aa}\\
&=&V^{-1} e^{-\beta m} \left[L(\xx)\right]_{aa}\;,
	\label{eq:quarkPropagator}
\eqa
where $V$ is the volume and 
with no sum over the color index $a$. One can take the vanishing
of the propagator and therefore the Polyakov loop as a sign of 
confinement. 

In the context of the PQM and $\chi M$ models, 
it is convenient to choose a gauge which simplifies the
Polyakov loop as much as possible. The Weyl gauge $A_4=0$
would make the Polyakov loop trivial; however this gauge is not compatible
with the periodicity requirement of the gauge field in the
imaginary time formalism \cite{Weiss1,Weyl1}.
Instead one can choose the so-called static gauge \cite{nadkarni}, where
\bqa
\partial_{\tau}A_4&=&0\;.
\eqa
Furthermore, one can rotate the gauge fields so that $A_4$
is in the Cartan subalgebra of the Lie algebra of $SU(N_c)$ \cite{skokovrev}.
In the case of $N_c=3$, the gauge field in the Polyakov gauge
can be written as
\bqa
A_4 = \frac{ 1 }{ 2 }(\lambda_3 A_4^3 + \lambda_8 A_4^8)\;,
\eqa
where $\lambda_3$ and $\lambda_8$ are the two diagonal Gell-Mann matrices.
Defining
\begin{equation}
q = \frac{ 3 }{ 4\pi } g\beta A_4^3\ , \quad r = \frac{ \sqrt{3} }{ 4\pi }
g\beta A_4^{8}\ ,
\end{equation}
we can express the background gauge field as
\begin{equation}
g\beta A_4 = \frac{ 2\pi }{ 3 }\diag(q+r, - q + r, -2r)\;.
\end{equation}
The thermal Wilson line can then be written as
\begin{equation}
L(\xx) = \begin{pmatrix}
e^{i\frac{ 2\pi }{ 3 }\left[q(\xx) + r(\xx)\right]} & 0 & 0 \\
0 & e^{i\frac{ 2\pi }{ 3 }\left[-q(\xx) + r(\xx)\right]} & 0 \\
	0 & 0 & e^{i \frac{ 2\pi }{ 3 }\left[-2r(\xx)\right]}
\end{pmatrix}\;.
\label{eq:polyakovMatrix}
\end{equation}
Taking the trace to obtain the Polyakov loop and its conjugate yields
\bqa
\label{phiex}
\Phi &=& \frac{ e^{2 \pi i r / 3 } }{ 3 }
\left[e^{-2\pi i r} + 2 \cos\left(\frac{ 2\pi q }{ 3 }\right)\right]\;,\\
\bar{\Phi} &=& \frac{ e^{-2 \pi i r / 3 } }{ 3 }
\left[e^{2\pi i r} + 2 \cos\left(\frac{ 2\pi q }{ 3 }\right)\right]\;.
\label{barphiex}
\eqa
When $A_4$ is constant in space, and thus also $r$ and $q$, we see that
\begin{equation}
\Phi =  
\begin{cases}
0, & q=1,\ r=0 \\
1, & q=0, \ r=0
\end{cases},
\end{equation}
at the classical level. Thus 
we conclude that 
a state with $q = 1,\ r = 0$ is a deconfined state, while a state with 
$q = 0,\ r = 0$ is a confined state.

In QCD we must have that $\big<\Phi\big>^* = \big<\Phi^\dag\big>$ for $\mu=0$, while 
for $\mu \neq 0$ it turns out that $\big<\Phi\big>^* \neq \big<\Phi^\dag\big>$ 
\cite{skokovrev,RRTW_finiteMu,Pisarski_finiteMu,QCD_spinModel_finiteMu}. 
Furthermore, it is found that $\big<\Phi\big>$ and $\big<\Phi^\dag\big>$ are 
both real, but with $\big<\Phi^\dag\big>\neq\big<\Phi\big>$ for $\mu\neq 0$ 
\cite{skokovrev, lattice_finiteMu, QCD_spinModel_finiteMu}. 
Why this must be the case is shown non-perturbatively in Refs.~\cite{ ComplexA4_1, skokovrev}.

\section{Gluonic sector}
\label{sec:gluon}
In this section, we discuss the gluonic sector of the effective(/grand) potential $\Omega$ of the PQM and $\chi M$ models, which is the main difference between the two models. 
They are somewhat different, but they both involve a phenomenological
pure-glue potential with a few parameters that are determined such that 
several physical quantities from pure glue lattice simulations are reproduced.

\subsection{PQM model}

It is known from lattice simulations that a first-order phase transition, 
corresponding to gluonic deconfinement, happens at $T_0=\SI{270}{\mega\electronvolt}$ 
in pure $SU(3)$ gauge theory \cite{PureGaugeTemp}. 
A first-order transition is what is expected on the basis of universality, 
as argued by Svetitsky and Yaffe in Refs. 
\cite{SvetitskyYaffe1,SvetitskyYaffe2}.
In addition to the knowledge of the location of the phase transition, various 
thermodynamical properties such as the pressure and internal energy as function 
of temperature have been established 
\cite{Boyd, YangMillsState1, YangMillsState2}. 
Finally, one also has simulations of the value of Polyakov loop as function of 
temperature \cite{PolyakovLattice}. 

With the knowledge of for example
$T_0$, $P(T)$ and $\Phi(T)$ from lattice simulations, one can 
write down a phenomenological potential $\mathcal{U}_{\mathrm{glue}}(\Phi, \bar{\Phi}, T)$
that reproduces these three quantities.
The first requirement necessitates that the form of the effective potential 
admits a first-order transition in the first place. 
For the second criterion, 
we can find the pressure from the effective potential as
$P=-\mathcal{U}_{\rm glue}(\Phi(T), \bar{\Phi}(T),T)$ with $\Phi$ and $\bar{\Phi}$ evaluated at the minimum of $\mathcal{U}_{\rm glue}$. 
Regarding the form of the potential, 
several things can be said on general grounds. 
It must be symmetric under center transformations since the gluonic action is 
center symmetric. Remembering the transformation rule for 
$\Phi$ and $\bar{\Phi}$ under center transformations, 
we see that the potential can be a function the terms 
$\Phi\bar{\Phi}$, $\Phi^3$ and $\bar{\Phi}^3$ only. Additionally, there is no 
reason for any asymmetry between $\Phi$ and $\bar{\Phi}$ in a pure gluonic 
system, and we thus require that the potential is symmetric under 
$\Phi \leftrightarrow \bar{\Phi}$. Finally, we should demand that the minimum of 
$\mathcal{U}_{\mathrm{glue}}$ at low temperatures is at $\Phi = \bar{\Phi} = 0$, while at 
high temperatures it should equal or asymptotically approach 
$\Phi = \bar{\Phi} = 1$.

Several potentials have been suggested in the literature 
\cite{Ratti2006, Ratti2007, Fukushima2008, Dexheimer2010}, and some of the more 
frequently used are compared in Ref. \cite{Schaefer2010}. 
The number of fit parameters vary from two \cite{Fukushima2008} to 
seven \cite{Ratti2006}. One of the models by Ratti, R{\"o}{\ss}ner, Thaler and 
Weise \cite{Ratti2007}, which is the one we use for the PQM model, takes the form
\begin{multline}
\frac{ \mathcal{U_\mathrm{RRTW}} }{ T^4 } = 
b(T) \ln\Big[1 - 6\Phi\bar{\Phi}
+4(\Phi^3 + \bar{\Phi}^3) 
- 3(\Phi\bar{\Phi})^2 \Big] \\
-\frac{ 1 }{ 2 } a(T) \Phi \bar{\Phi}\ , 
\label{eq:RRTW_pot_repeat}
\end{multline}
with the temperature-dependent coefficients
\begin{align}
	a(T) &= a_1 + a_2 \left(\frac{ T_0 }{ T }\right) + a_3 \left(\frac{ T_0 }
{ T }\right)^2, \\
	b(T) &= b_1 \left(\frac{ T_0 }{ T }\right)^3.
\end{align}
We take all the parameters except $T_0$ as given in the original paper, meaning
\begin{equation}
a_1 = 3.51\ ,\quad a_2 = -2.47\ ,\quad a_3 = 15.2\ ,\quad b_1 = -1.75\ . 
\label{eq:RRTW_pars}
\end{equation}
We note that the potential presented above does not take into account the
backreaction of quarks onto the gluonic sector. However, from the fact that the running coupling in QCD  depends
on the number of quark flavors, as evident from the one-loop expression
\begin{equation}
	\frac{ g^2(\Lambda, N_f) }{ 4\pi } = \frac{ 2\pi }{ (11 - \frac{ 2 }{ 3 }N_f) 
}\frac{ 1 }{ \ln(\frac{ \Lambda }{ \Lambda_{\mathrm{QCD}} }) }\ ,\label{eq:QCDrunning}
\end{equation}
it is natural to let $T_0=T_0(N_f)$, so that the behavior in the gluonic sector also depends on 
$N_f$, since $g$ determines the strength of the interactions between the gauge 
fields. 
The authors of Ref. \cite{Schaefer2007} 
parametrize this dependence as
\begin{equation}
	T_0(N_f) = \hat{T}e^{- \frac{ 2\pi }{ \alpha_0 }(11-\frac{ 2 }{ 3 }N_f)^{-1}}\;,
	\label{eq:ScheaferT0}
\end{equation}
where the constants are
$\hat{T} = \SI{1.77}{\giga\electronvolt}$ and
$\alpha_0 = 0.304$.
This expression is heuristically obtained by assuming that the temperature 
dependence of $g$ is governed by \eqref{eq:QCDrunning} with $\Lambda = T$ and that 
the deconfinement phase transition occurs at a specific coupling, so that we can solve
\begin{equation}
	g(T_0(N_f), N_f) = g(T_0(0), 0)
\end{equation}
for $T_0(N_f)$. With $T_0(N_f = 0) = \SI{270}{\mega\electronvolt}$ 
we get $T_0(N_f = 2) = \SI{208}{\mega\electronvolt}$.

\subsection{Chiral matrix model}
The gluonic part of the chiral matrix model was developed as an effective
model for pure $SU(3)$ gauge theory in Refs. \cite{PureSU3_MM1,PureSU3_MM2},
where the degrees of freedom are $r$ and $q$ only. The potential
consists
of two terms; one is obtained by integrating out a fluctuating gauge field in a background gauge field $A_4$ to one loop. The other term models a nonperturbative
contribution and is added by hand. 

The one-loop perturbative contribution to the effective potential of pure 
$SU(3)$ Yang-Mills theory reads 
\bqa
	\mathcal{V}_{\mathrm{pt}}(q, r) = \pi^2 T^4 \left[-\frac{ 8 }{ 45 }+
\frac{ 4 }{ 3 }\mathcal{V}_2(q, r) \right]\;,
\label{eq:vptchim}
\eqa
where
\bqa
\mathcal{V}_2(q, r) &= B_2\left(\frac{ 2q }{ 3 }\right) + 
B_2\left(\frac{ q }{ 3 } + r\right) + B_2\left(\frac{ q }{ 3 } - r\right)
\label{eq:V1chiM}
\eqa 
and 
\bqa
B_k(x) &= \left| x \right|_{\mathrm{mod } 1}^k(1-|x|_{\mathrm{mod } 1})^k\ .
\eqa
The second term in Eq.~\eqref{eq:vptchim} is the Weiss potential, first calculated by
Weiss \cite{Weiss1, Weiss2} and Gross, Pisarski, and Yaffe \cite{GrossPisarskiYaffe}.
To drive the system to confinement at low temperatures one adds a 
phenomenological potential, which is chosen to be of the form 
\bqa\nonumber
\mathcal{V}_{\mathrm{nonpt}}(q, r) &=&
 -\frac{ 4\pi^2 }{ 3 } T^2 T_d^2 \left[\frac{ c_1 }{ 5 } \mathcal{V}_1(q,r) + 
c_2 \mathcal{V}_2(q, r) 
\right.\\
&&\left.
- \frac{ 2 }{ 15 }c_3 \right]\;,
	\label{eq:nonpt-pot}
\eqa
where  
\bqa
\mathcal{V}_1(q, r) &= B_1\left(\frac{ 2q }{ 3 }\right) + 
B_1\left(\frac{ q }{ 3 } + r\right) + B_1\left(\frac{ q }{ 3 } - r\right), \label{eq:V2chiM}
\eqa
with four fit parameters $c_1$, $c_2$, $c_3$ and $T_d$. At temperatures below roughly $T\sim T_d$ 
the $\sim T^2$ term will dominate over the $\sim T^4$ perturbative term, 
and it can thus drive the system to confinement with an appropriate choice of 
the fit parameters. The $T^2$ behavior is chosen since it has been observed 
in lattice data that the subleading contribution to the pressure goes as 
$\sim T^2$ \cite{T2_pressure1, T2_pressure2}. 
The parameters $c_1$ and $c_3$ are chosen so that the pressure in the confined 
phase of the pure gauge theory is zero and so that a phase transition happens 
at $T_d$. The former is an approximation, but it is reasonable since 
the pressure of the confined phase in $SU(3)$ gauge theory is very 
low compared to the deconfined phase, as lattice data show \cite{Boyd}. 
Furthermore, we choose $T_d = \SI{270}{\mega\electronvolt}$, which 
is roughly the deconfinement temperature in $SU(3)$ gauge theory \cite{Boyd}. 
Then only $c_2$ remains as a fit parameter. It is determined by fitting the 
interaction measure $(\mathcal{E}-3P)/T^4$ predicted by the full gluonic 
potential,
\begin{equation}
\mathcal{U}_{\chi M} = \mathcal{V}_{\mathrm{pt}} + \mathcal{V}_{\mathrm{nonpt}}\;,
\end{equation}
to lattice data, and the result is $c_2 = 0.830$ \cite{PisarskiSkokov}, which 
gives 
\begin{equation}
c_1 = 0.315, \quad c_2 = 0.830, \quad c_3 = 1.13\;. 
\end{equation}

\begin{figure*}[htb!]
    \includegraphics[width=1\textwidth]{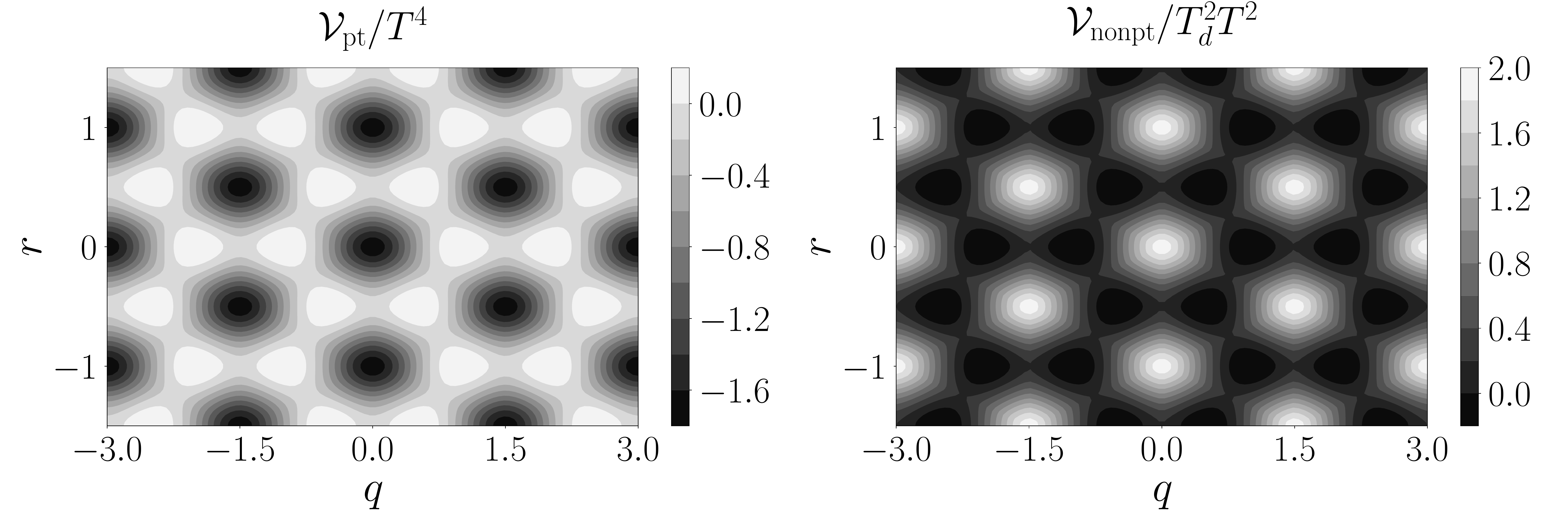}
    \caption{Contour plots of the perturbative (left) and nonperturbative (right) contributions to the
    gluonic potential in the $\chi M$ model.}
        \label{fig:gluonPot2}
\end{figure*}

In Fig. \ref{fig:gluonPot2}, we show contour plots of 
the perturbative (left panel)
and the nonperturbative (right panel) contributions to the gluonic
potential in the chiral matrix model. 
The perturbative
contribution ${\cal V}_{\rm pt}(q,r)$ has minima at
$q=r=0$, $q=0$ and $r=1$ etc., and maxima at $q=1$, $r=0$ and $q=r={1\over2}$
etc. The potential reflects the center symmetry of $SU(3)$.
The nonperturbative potential ${\cal V}_{\rm nonpt}(q,r)$ behaves the opposite
way; it has minima where the perturbative potential has maxima and
vice versa. This potential also reflects the center symmetry.
The full gluonic potential is then the sum of the two contributions,
and the latter has been constructed so that the two terms are competing.
At high temperatures $\mathcal{V}_{\mathrm{pt}}$ 
dominates while for low temperatures $\mathcal{V}_{\mathrm{nonpt}}$ dominates. 

\section{Chiral sector}
\label{sec:chiral}
In this section, we will discuss the chiral sector of the two models.
In Ref.~\cite{PisarskiSkokov}, Pisarski and Skokov add a phenomenological 
quark term to the $\chi M$ model which is not present in the QM model, but apart from this the chiral sector in the \chiM{} model corresponds to the QM model.
Adding this phenomenological term to the PQM as well, their chiral sectors are identical. Note however that while Ref.~\cite{PisarskiSkokov} includes the strange quark, we treat only the two light quark flavors.

\subsection{Quark-meson model}
To obtain the two-flavor quark-meson model we couple 
two $N_c$-plets of fermionic fields via Yukawa interactions to the linear
sigma model with an approximate $SU(2)_L\times SU(2)_R$ symmetry.
The fields $\psi_1$ and $\psi_2$ are taken to represent up and down quarks, 
respectively.
The Lagrangian of the two-flavor quark-meson model in Minkowski space is 
\bqa\nonumber
{\cal L}&=&
{1\over2}\left[(\partial_{\mu}\tilde{\sigma})(\partial^{\mu}\tilde{\sigma})
+(\partial_{\mu} \pi_3)(\partial^{\mu} \pi_3)
\right]
\\&&\nonumber
+(\partial_{\mu}+2i\mu_I\delta_{\mu}^0)\pi^+(\partial^{\mu}-2i\mu_I\delta_{0}^{\mu})
\pi^-
\\&&\nonumber-{1\over2}m^2(\tilde{\sigma}^2+\pi_3^2+2\pi^+\pi^-)
-{\lambda\over24}(\tilde{\sigma}^2+\pi_3^2+2\pi^+\pi^-)^2
\\ && \nonumber
+h\tilde{\sigma}+\bar{\psi}\left[
i/\!\!\!\partial
+ \mu
\gamma^0
-g(\tilde{\sigma}+i\gamma^5{\boldsymbol\tau}\cdot{\boldsymbol\pi})\right]\psi\;,
\\ &&
\label{eq:QMlag}
\eqa
where $\psi$ is the flavor doublet 
\begin{align}
\psi =&  \begin{pmatrix}
           \psi_1 \\
           \psi_2
         \end{pmatrix},
\end{align}
and $\boldsymbol{\pi}$ is the isospin triplet $(\pi_1, \pi_2, \pi_3)^T$, with $\pi^{\pm} = (\pi_1 \pm i\pi_2)/\sqrt{2}$. $\tilde{\sigma}$ is a scalar isospin singlet that will attain a vacuum expectation value that corresponds to the chiral condensate.
The  $\tau_i$ are the Pauli matrices acting in flavor space and 
$\gamma^5 \equiv i \gamma^0 \gamma^1 \gamma^2 \gamma^3$.
$\mu_I$ is the isospin chemical potential and $\mu = \frac{1}{3}\mu_B$ the quark chemical potential. 

Let us identify the symmetries of the QM model. In the chiral limit, meaning $h=0$, the QM 
Lagrangian has a global $SU(N_c) \times SU(2)_L \times SU(2)_R \times U(1)_B$ 
symmetry, while at the physical point ($h\neq 0$) the symmetry is 
$SU(N_c) \times SU(2)_V \times U(1)_B$. The $U(1)_B$ symmetry gives rise to conservation of baryon number, with the associated baryon chemical potential $\mu_B={3\over2}(\mu_u+\mu_d)=3\mu$,
where $\mu_u$ and $\mu_d$ are the up and down quark chemical potentials, respectively.
The isospin chemical potential $\mu_I$ is given by $\mu_I={1\over2}(\mu_u-\mu_d)$.
When $\mu_I\neq0$, i.e. when
$\mu_u\neq\mu_d$, the $SU(2)_V$ is reduced to $U(1)_{I_3L}\times U(1)_{I_3R}$ for $h=0$
and $U(1)_{I_3}$ if $h\neq0$.

The full chiral sector of the PQM and \chiM{} models is obtained by coupling the quarks to a temporal gauge field in the Euclidean Lagrangian, 
\begin{equation}
    \tilde{\gamma}^{\mu} \partial_\mu \rightarrow\tilde{\gamma}^{\mu} (\partial_\mu - ig_{\mathrm{YM}} \delta_{\mu}^4 A_4 )\ ,
\end{equation}
where $\tilde{\gamma}_4 = \gamma^0$ and $\tilde{\gamma_i} = -i\gamma^i$ are the Euclidean gamma matrices, $\delta_{\mu\nu}$ the Euclidean metric, and $\mu \in \left\{1, 2, 3, 4 \right\}$. 

We take $m^2 < 0$ so that $\tilde{\sigma}$ attains a vacuum expectation value $v$, and define $\tilde{\sigma} = v + \sigma$. Here $v$ is the chiral condensate.
To obtain the effective potential $\mathcal{U}_{\mathrm{chiral}}$,
we work to one-loop order and neglect bosonic fluctuations. As mentioned, the latter approximation is equivalent to taking the large-$N_c$ limit. 
The contribution to the thermodynamic potential from the Lagrangian \eqref{eq:QMlag} coupled to the background gauge field  then consists of two terms; a vacuum term arising from the tree-level
mesonic potential and the fermion determinant, and a thermal
piece coming from the same fermion determinant. 

For $\mu_I=0$ and with $\Delta \equiv gv$, we write
\bqa\nonumber
	\mathcal{U}_{\mathrm{chiral}}(\Delta, q, r, T, \mu) &=& 
        \mathcal{U}_{ \mathrm{vac}}(\Delta)+ \nonumber
        \mathcal{U}_{q, T}(\Delta, q, r, T, \mu)
\label{eq:renormGrandPot_RG}
\eqa
where we have, to one-loop order after renormalization and consistent parameter fixing,
\begin{widetext}
\bqa
\mathcal{U}_{\mathrm{vac}}(\Delta)&=& \frac{ 3 }{ 4 }m_{\pi}^2 f_{\pi}^2
\left\{1 - \frac{ 4 N_c m_q^2 }{ (4\pi)^2 f_{\pi}^2 }m_\pi^2 F^{\prime}(m_{\pi}^2)
\right\} 
\frac{ \Delta^2 }{ m_q^2 } + \frac{ 2N_c m_q^4 }{ (4\pi)^2 }\left(\frac{3}{2} - 
\ln\frac{\Delta^2}{m_q^2} \right)\frac{ \Delta^4 }{ m_q^4 }  \nonumber \\ 
&& - \frac{m_\sigma^2 f_{\pi}^2 }{ 4 }
\left\{1 + \frac{ 4 N_c m_q^2 }{ (4\pi)^2 f_{\pi}^2  }\left[\left(1-\frac{ 4m_q^2}
{ m_\sigma^2 }\right)F(m_\sigma^2) 
- F(m_{\pi}^2) - m_\pi^2F^{\prime}(m_\pi^2) + \frac{ 4m_q^2 }{ m_\sigma^2 }  \right]
\right\} 
\frac{\Delta^2}{m_q^2} \nonumber \\ \nonumber
&& + \frac{ m_{\sigma}^2 f_\pi^2}{ 8 } \left\{1 + \frac{ 4N_c m_q^2 }
     { (4\pi)^2 f_\pi^2 }\left[ \left(1 - \frac{ 4m_q^2 }{m_\sigma^2 }\right)
       F(m_\sigma^2) - 
       F(m_{\pi}^2) - m_{\pi}^2F^{\prime}(m_{\pi}^2)\right]
     \right\}\frac{ \Delta^4 }{ m_q^4 } 
\\
&&- \frac{ m_\pi^2 f_\pi^2 }{ 8 }\left\{ 1 - \frac{ 4N_c m_q^2 }
{ (4\pi)^2 f_{\pi}^2 } m_\pi^2 F^{\prime}(m_{\pi}^2) \right\} \frac{\Delta^4}{m_q^4} 
- m_\pi^2 f_\pi^2\left[1-\frac{4N_c m_q^2}{(4\pi)^2 f_\pi^2}m_\pi^2F^{\prime}(m_{\pi}^2)
  \right]
\frac{\Delta}{m_q}\;,
\label{eq:renormGrandPot}
\\
\nonumber
	\mathcal{U}_{q, T}(\Delta, T, \mu, q, r) &=& 
-4 T \ddpp \bigg\{\tr_c \ln\left[1 + L e^{-\beta(\omega_{\pp} - \mu)}\right] + 
\tr_c \ln\left[1 + \bar{L} e^{-\beta(\omega_{\pp} + \mu)}\right]\bigg\}
\\ &=&
- 4 T  \ddpp \ln\left[1 + 3\Phi e^{-\beta(\omega_{\pp}-\mu)}+ 
3\bar{\Phi} e^{-2 \beta(\omega_{\pp}-\mu)} + e^{-3\beta(\omega_{\pp}-\mu)}\right] \nonumber 
\\
	      & &- 4 T \ddpp \ln\left[1 + 3\bar{\Phi} e^{-\beta(\omega_{\pp}+\mu)}+ 
3\Phi e^{-2 \beta(\omega_{\pp}+\mu)} + e^{-3\beta(\omega_{\pp}+\mu)}\right]\;.
\label{eq:quarkTfluct}
\eqa
\end{widetext}

Here $m_q$, $m_\pi$ and $m_\sigma$ are the physical quark, pion and sigma masses at $T=0$, respectively, while $f_\pi$ is the pion decay constant. The quantity $\Delta$ is, in addition to the rescaled chiral condensate, the constituent quark mass, and satisfies $\Delta(T=\mu=0)=m_q$.
The derivation of Eq.~(\ref{eq:renormGrandPot}) can be found in Appendices
A and B, and the definition of the functions $F(m^2)$ and
$F^{\prime}(m^2)$ are given in Eqs.~(\ref{fdef}) and (\ref{fpdef}).
Equation \eqref{eq:quarkTfluct} is obtained in the same way as one would calculate the free fermion partition function, except one now has a complex effective chemical potential $\tilde{\mu}_j$ that differs for each quark color $j$, with
\begin{equation}
   \tilde{\mu}_j = \mu + ig[A_4]_{jj}\ ,
\end{equation}
with $[A_4]_{jj}$ the $j$-th diagonal element. 
This derivation requires using the Polyakov gauge, where $A_4$ is diagonal.

\begin{figure}
	\includegraphics[width=0.4\textwidth]{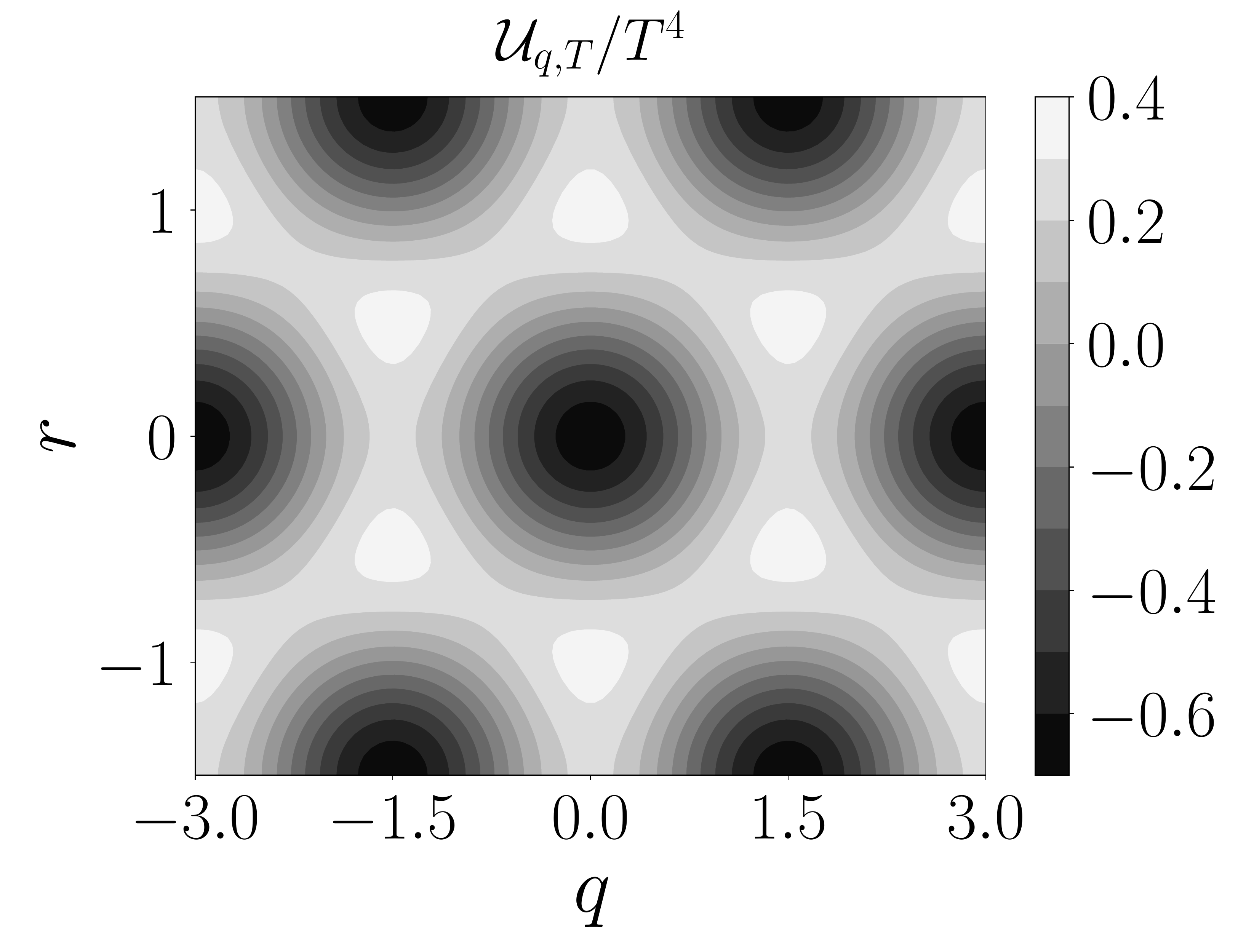}
\caption{Contour plot of 
$\mathcal{U}_{q,  T}(q, r)$ for $\mu = 0$, 
$\Delta = \SI{300}{\mega\electronvolt}$, and $T=\SI{100}{\mega\electronvolt}$.}
	\label{fig:quarkThermalPot}
\end{figure}

The thermal quark potential $\mathcal{U}_{q, T}$  as function of $q$ and $r$
at $\mu=0$, $T=\SI{100}{\mega\electronvolt}$, and $\Delta = \SI{300}{\mega\electronvolt}$ is shown 
in Fig.~\ref{fig:quarkThermalPot}. 
We see that this potential, whose qualitative shape is mostly unchanged for 
other values of $\Delta$ or $T$, drives $(q,r)$ towards $q=r=0$, which 
corresponds to deconfinement. Thus, it is expected that the addition of 
quarks to the gluonic potential lowers the deconfinement temperature.

We finally note that for $\mu \neq 0$, Eq.~\eqref{eq:quarkTfluct} can become complex. This will be discussed in Sec.~\ref{sec:minProcedure}.

\subsection{A phenomenological quark term}
\label{chimodel}
In addition to the (partly) phenomenological gluonic sector, Pisarski and 
Skokov add to the \chiM{} model a phenomenological quark term. In the two-flavor case, it is given by 
\begin{equation}
	\mathcal{U}_{q, \mathrm{cur}}(\Delta, q, r, T, \mu) = - m_{\mathrm{cur}} 
\frac{ \partial }{ \partial \Delta } \mathcal{U}_{q, T}\ ,
\end{equation}
where $m_{\mathrm{cur}}$ is the current quark mass. 
This term is added in order to achieve that 
$\Delta \rightarrow m_{\mathrm{cur}}$ in the high-temperature limit. 
Let us show how this works: In the high-temperature limit we 
expect $q=r=0$, and we can thus set the Polyakov loop to be 
$\Phi=\bar{\Phi}=1$. Let us furthermore assume that $\mu=0$ and
$T \gg \Delta$. 
We can expand $\mathcal{U}_{q, T}$ for $\mu=0$, $\Phi=\bar{\Phi}=1$ in 
powers of ${\Delta\over T}$ as 
\bqa\nonumber
\frac{\mathcal{U}_{q, T}}{N_f N_c}&=& -4 T \ddpp \ln\left[1 + e^{-\beta\omega_{\pp}}\right] 
\\
&\approx& - \frac{ 7\pi^2 }{ 180 }T^4 + \frac{T^2  \Delta^2  }{ 12 } + 
\mathcal{O}
\left(\Delta^4 \ln\frac{ \Delta }{ T }\right).
\eqa
Thus, to leading order we find 
\begin{equation}
	\mathcal{U}_{q, \mathrm{cur}}/N_f N_c = - \frac{ 1 }{ 6 } m_{\mathrm{cur}} T^2 
\Delta\;.
\end{equation}
As we assume high temperatures, we consider the potential only up to subleading 
temperature dependence $\sim T^2$. Furthermore, we assume that $\Delta$ is 
small, which we expect in the high-temperature phase where chiral symmetry is 
approximately restored. Using this we keep only leading and subleading terms in 
$\Delta$. Thus, in the high temperature limit we find that the effective potential goes as 
\begin{equation}
\frac{ \Omega }{ N_f N_c } \approx - \frac{ 7\pi^2 }{180 }T^4 + 
\frac{ T^2 \Delta^2}{ 12 }- \frac{ 1 }{ 6 } m_{\mathrm{cur}}T^2\Delta 
+ 
\mathcal{O}\left(\Delta^4 \ln{\frac{ \Delta }{ T }}\right)\;.
\end{equation}
Minimizing this potential with respect to $\Delta$, we immediately find 
\begin{equation}
\Delta = m_{\mathrm{cur}}\ ,
\end{equation}
and we expect $\Delta \rightarrow m_{\mathrm{cur}}$ in the high-temperature limit. 

When we later in this section investigate the thermodynamics of the PQM and 
\chiM{} models, we will assess the effects of $\mathcal{U}_{q, \mathrm{cur}}$ on 
the thermodynamic functions. Due to its ad hoc nature it should preferably affect the 
thermodynamics minimally while still achieving its purpose of ensuring the 
quark mass to approach the current quark mass in the approximately 
chirally restored phase.

\section{Minimizing $\Omega$ at $\mu \neq 0$}
\label{sec:minProcedure}

There is one more problem we must face.
For the case of $\mu=0$, the quark effective potential
is real for any $q, r \in \mathbb{R}$ since the two terms in Eq.~\eqref{eq:quarkTfluct} are complex conjugates of each other. 
However, upon introducing of $\mu \neq 0$ 
this breaks down, and the potential
becomes complex in general. 

The solution suggested in Refs. \cite{ComplexA4_1, ComplexA4_2, PisarskiSkokov} is, when $\mu\neq0 $, to let the background $A_4$ field become non-Hermitian by
setting
$q\in \mathbb{R}$ and $r=iR$ with $R\in \mathbb{R}$. 
The rationale behind taking $r$ imaginary is discussed in Refs.~\cite{ComplexA4_1, ComplexA4_3}.
It is not as unreasonable as it first seems, since $A_4$ in the PQM and \chiM{} models represents the mean field of a quantum field, $A_4 = \expect{A_4^{\mathrm{qu}}}$, and when $\mu \neq 0 $ in full QCD the Euclidean action becomes complex. Because of this, even if each field configuration is real, when carrying out the path integral where we weight each field configuration with $e^{-S_E}$, we might get that $\expect{A_4^{\mathrm{qu}}}$ is complex.

\begin{figure}[htb!]
	\centering
		\centering
\includegraphics[width=0.4\textwidth]{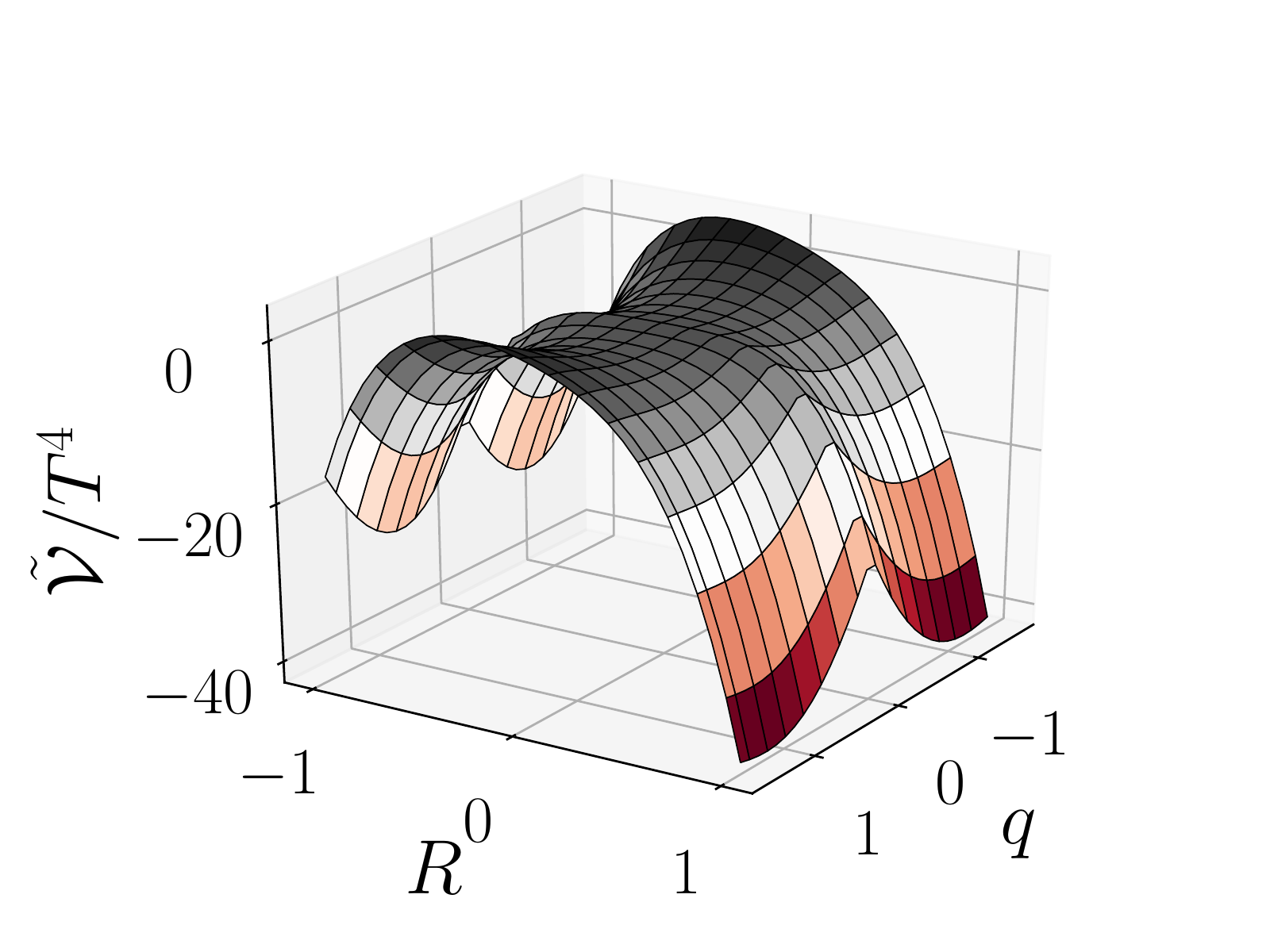}
\caption{Normalized $R$-dependent part of the effective potential in the 
$\chi M$ model at $T=\mu=200$ MeV, and $\Delta= 150$ MeV.}
		\label{fig:saddle}
\end{figure}

Inserting $r=iR$ into Eqs. (\ref{phiex}) and (\ref{barphiex}), we find
\begin{align}
\Phi &= \frac{ e^{- 2 \pi R / 3 } }{ 3 }\left[e^{2\pi R} + 2 
\cos\left(\frac{ 2\pi q }{ 3 }\right)\right]\;, \\
\bar{\Phi} &= \frac{ e^{2 \pi R / 3 } }{ 3 }\left[e^{-2\pi R} + 2 
\cos\left(\frac{ 2\pi q }{ 3 }\right)\right]\;, 
\end{align}
which gives that both are real, but with different values. For the RRTW 
potential, which is a function of $\Phi$ and $\bar{\Phi}$, it is clear that the 
Polyakov-loop potential becomes real, and thus the full potential is also real 
for all $(q, R) \in \mathbb{R}^2$. For the \chiM{} model this is also the case, 
since the only potentially complex terms in Eqs.~\eqref{eq:V1chiM} and 
\eqref{eq:V2chiM} are
\begin{equation}
B_1\left(\frac{ q }{ 3 }+iR\right) + B_1\left(\frac{ q }{ 3 }-iR\right) = 
2\Re B_1\left(\frac{ q }{ 3 }-iR\right),
\end{equation}
and 
\begin{equation}
B_2\left(\frac{ q }{ 3 }+iR\right) + B_2\left(\frac{ q }{ 3 }-iR\right) = 
2\Re B_2\left(\frac{ q }{ 3 }-iR\right).
\end{equation}
However, the full effective potential $\Omega_{\chi M}$
is unbounded as a function of $R$ for low temperatures, and
$\Omega_{\rm PQM}$
is unbounded as a function of $R$ for all temperatures.
The part of the potential that depends on
$R$ in the $\chi M$ model,
$\tilde{V}={\cal U}_{q,T}+{\cal U}_{\chi M}$, is shown in Fig.~\ref{fig:saddle}
as function of
$(q,R)$. We see that
there is no minimum for $|R|<1$. This behavior persists for any
$R$.
In Refs. \cite{ComplexA4_1, ComplexA4_2} the authors deal with this by arguing that the physically realized state is a saddle point
of $\Omega(q,R)$. This approach gives
$\Phi\neq\bar{\Phi}$ with both being real, thus giving a real effective potential. That we obtain $\Phi\neq\bar{\Phi}$ is desirable, since we do not expect the free energy of single a quark to be equal to that of a single antiquark when $\mu\neq0$. 
However, in setting $r$ imaginary and choosing a saddle point some ambiguity remains with respect to what saddle point to choose and why $r$ is purely imaginary and $q$ purely real. In Ref.~\cite{ComplexA4_1} the saddle point with the lowest energy is chosen.
It is pointed out in Ref.~\cite{Re_vs_saddle} that it is not known if interface tensions can be calculated within this scheme.

An alternative approach, which is the one used in the following, is to keep $q, r \in \mathbb{R}$ and minimize $\Re \Omega$ under the constraint $\Im \Omega = 0$. If we interpret a complex $\Omega$ as signaling an unstable state, this might be reasonable. It turns out that a global minimum of $\Re \Omega$ can always be found at $r=0$, and with $r=0$ we always have  $\Im \Omega = 0$. This means that we can set $r=0$ and minimize $\Omega$ freely with respect to $q$ only. However, this scheme gives  $\Phi=\bar{\Phi} \in \mathbb{R}$, which is not what we expect from the quark/antiquark free energy interpretation of $\Phi$ and $\bar{\Phi}$. 
However, the equality of the two can be seen as a result of the fact that we are doing a mean-field treatment of $A_4$ instead of a mean-field treatment of the actual Polyakov loops. The quantities we are calling $\Phi$ and $\bar{\Phi}$ in the PQM model are, in the Polyakov gauge, 
\begin{equation}
    \Phi = \frac{1}{N_c} \tr_c e^{ig\expect{A_4^{\mathrm{qu}}}}, \quad \bar{\Phi} = \frac{1}{N_c} \tr_c e^{-ig\expect{A_4^{\mathrm{qu}}}},
    \end{equation}
which are not equivalent to the expectation value of the Polyakov loop quantum operators. Thus, the free energy interpretation should not be taken too seriously. It would however be useful to carry out a comparison between the two schemes in the future. 

\section{Thermodynamics}
\label{sec:thermo}

We now have all the ingredients needed to investigate the thermodynamics of the 
PQM and \chiM{} models at one loop in the large-$N_c$ limit. For a given 
temperature and chemical potential we numerically solve
\bqa
\quad \frac{ \partial \Omega }{ \partial q }= 0\;, 
\quad \frac{ \partial \Omega }{ \partial \Delta } = 0\;, 
\label{eq:potMinimize}
\eqa
with $r=0$ and require that we have a global minimum, where the full effective potential $\Omega$ for the two models reads
\begin{widetext}
\bqa
\Omega_{\chi M} &=&
\mathcal{U}_{\mathrm{vac}}(\Delta) + 
\mathcal{U}_{q, T}(\Delta, r, q, T, \mu) + 
\mathcal{U}_{q, \mathrm{cur}}(\Delta, r, q, T, \mu) 
+ \mathcal{U}_{\chi M}(r, q, T) - P_{0, \chi M}\;, 
	\label{eq:effPotChiM}
\\ 
\Omega_{\mathrm{PQM}} &=& \mathcal{U}_{ \mathrm{vac}}(\Delta) +
\mathcal{U}_{q, T}(\Delta, r, q, T, \mu) + 
\mathcal{U}_{q, \mathrm{cur}}(\Delta, r, q, T, \mu) 
+ \mathcal{U}_{\mathrm{RRTW}}(r, q, T) -P_{0, \mathrm{PQM}}\;. 
\label{eq:FinalEffPotPQM}
\eqa
\end{widetext}
We also add the term 
$\mathcal{U}_{q,\mathrm{cur}}$ to the PQM model, for the same reason that it is 
added to the \chiM{} model.
The parameters $P_{0, \chi M}$ and $P_{0, \mathrm{PQM}}$ are constants
that we subtract from the effective potential so that the condition
\begin{equation}
P(T=\mu=0) = 0\;,
\end{equation}
is satisfied for each of the two models. This constant will turn out to be small and has a negligible 
effect on the thermodynamics. However, it makes thermodynamic quantities 
divided by $T^4$ better behaved at temperatures close to zero. 

Once $\Delta$ and $q$ are determined as functions of $T$ and $\mu$, we can 
determine $\Omega$ as a function of $T$ and $\mu$ only. We can then calculate 
the pressure $P$, quark density $n_q=\expect{N}/V$, energy density 
$\mathcal{E}$ and interaction measure $I = (\mathcal{E}-3P)$ as functions of $\mu$ 
and $T$ via the relations 
\begin{align}
	P(T, \mu) &= - \Omega(\Delta(T), q(T), \mu)\;, \\
	n_q(T, \mu) &= \frac{ \partial P }{ \partial \mu }, \\
\mathcal{E}(T, \mu) &=  \mu n_q - P + T\frac{ \partial P }{ \partial T }\;.
\end{align}
To determine the one-loop couplings we use the following values for the masses
and the pion decay constant 
\bqa
	m_q &=& \SI{300}{\mega\electronvolt}\;, \\ 
	m_{\pi} &=& \SI{140}{\mega\electronvolt}\;,\\
	m_{\sigma} &=& \SI{500}{\mega\electronvolt}\;,\\
	f_{\pi} &=& \SI{93}{\mega\electronvolt}\;,\eqa
which yields the parameters
\bqa
	\lambda_0 &=& 61.5\ ,\\
	m_0^2 &=& -(\SI{449}{\mega\electronvolt})^2\ ,\\
	g_0 &=& 3.22\ ,\\
	h_0 &=& (\SI{121}{\mega\electronvolt})^3\ ,
\eqa
which are the one-loop values of the running couplings in the $\MS$ scheme at the renormalization scale 
\bqa\nonumber
\Lambda_0^2&=&m_q^2 \exp\left[- \Re F(m_\pi^2) - m_{\pi}^2
  \Re F^{\prime}(m_{\pi}^2)\right] 
\\
&=& (289\mathrm{MeV})^2\;.
\eqa
This is the scale that is consistent with $\expect{\sigma}$=0 in the on-shell scheme.

The sigma particle is a broad resonance whose mass is usually taken to be in the \SIrange{400}{800}{\mega\electronvolt} range, and the most recent estimated mass range is \SIrange{400}{550}{\mega\electronvolt} \cite{PDG}. We have chosen a value of \SI{500}{\mega\electronvolt}, but we vary it to
gauge the sensitivity of our results.

\subsection{Order parameters}

\begin{figure}[htb]
	\centering
		\centering
		\includegraphics[width=0.4\textwidth]{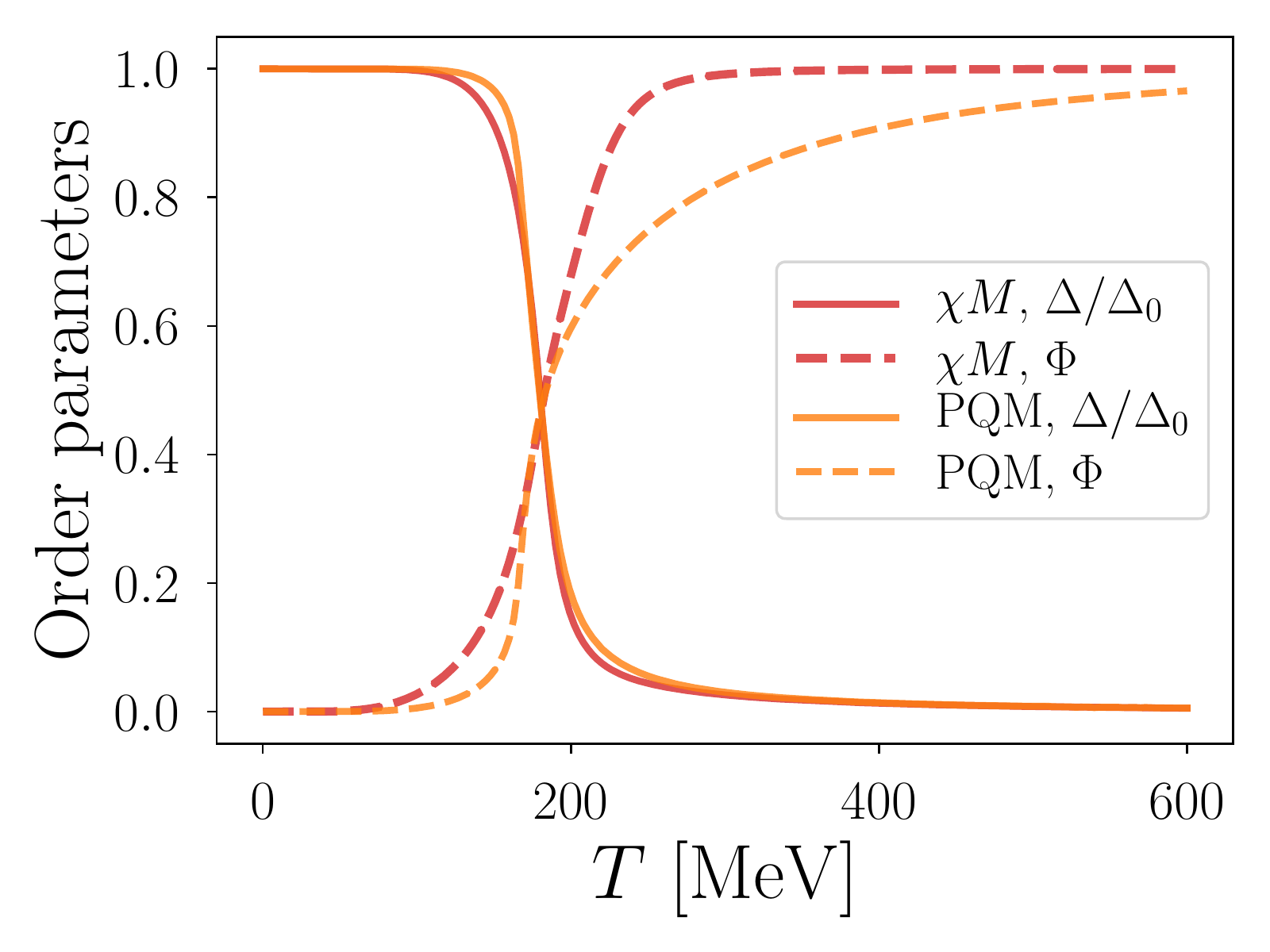}
		\caption{Order parameters
${\Delta(T)\over \Delta(T=0)}$ and $\Phi(T)$
                  in the \chiM{} and PQM models as 
function of temperature for $\mu=0$.}
		\label{fig:orderParams_a}
\end{figure}

\begin{figure}[htb]
		\centering
\includegraphics[width=0.4\textwidth]{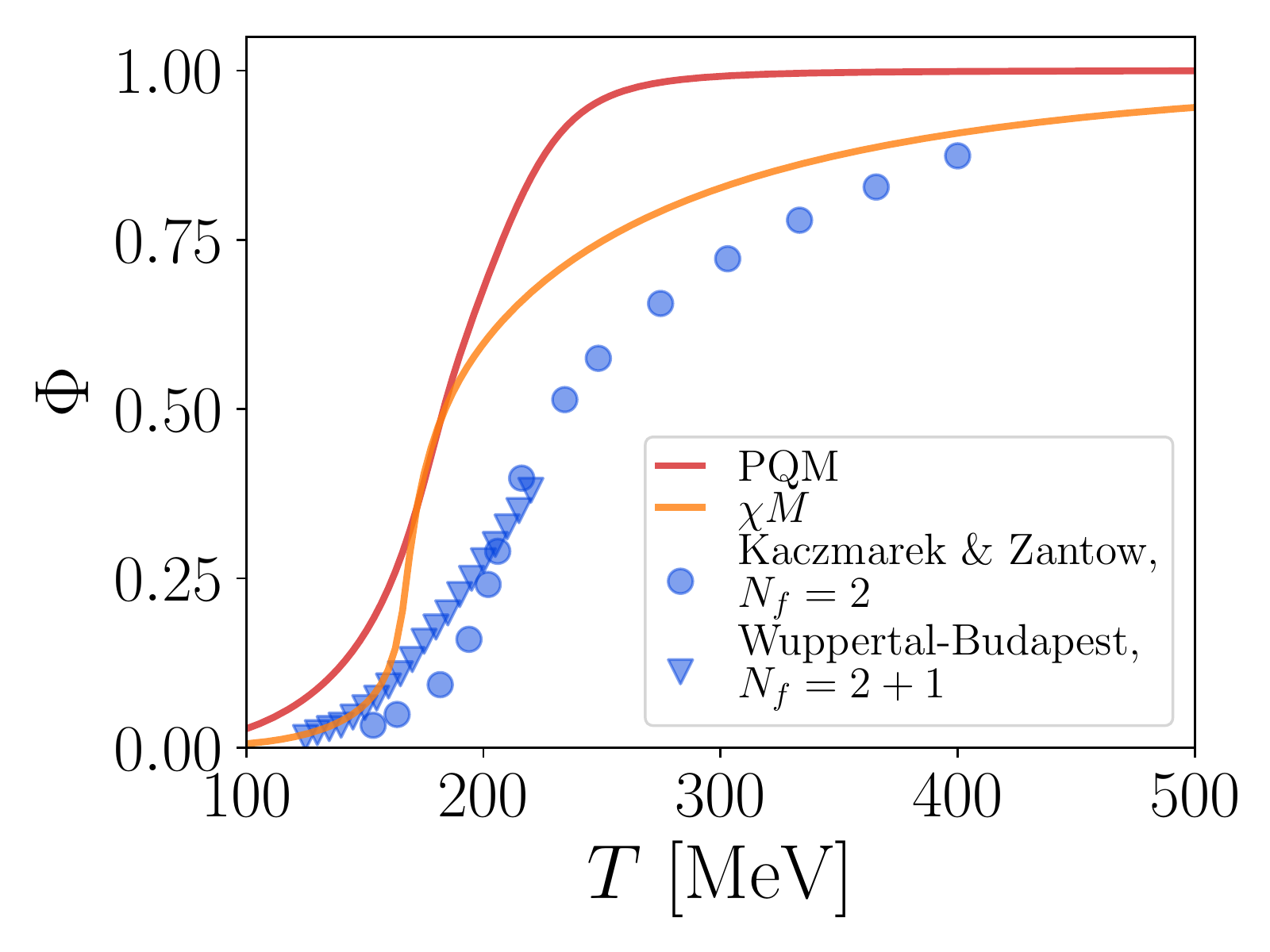}
\caption{Comparison of the Polyakov loop given by effective models and lattice calculations from Refs.~\cite{Wuppertal2010_Tc, KacZant_Polya}. }
		\label{fig:latticePolya}
\end{figure}

In Fig.~\ref{fig:orderParams_a}, we show the order parameters 
${\Delta(T)\over \Delta(T=0)}$ and $\Phi(T)$.
We point out that $\Delta$ does not go to zero at high temperatures, but 
rather approaches $\Delta\approx m_{\mathrm{cur}}$, 
as expected from the discussion in 
Sec.~\ref{chimodel}. 
We find that the \chiM{} model reaches full deconfinement 
at $T \approx 250{\rm MeV}$, while the PQM model reaches $\Phi=1$ more slowly, 
and it is in a ``semi-deconfined'' state between roughly 
200 and \SI{400}{\mega\electronvolt}. Like in Ref.~\cite{PisarskiSkokov} we find that the Polyakov loop in both models rises faster than on the lattice, which can be seen from Fig.~\ref{fig:latticePolya} where the models are compared to lattice data from Refs. \cite{Wuppertal2010_Tc, KacZant_Polya}. 

\begin{figure}[htb]
		\centering
\includegraphics[width=0.4\textwidth]{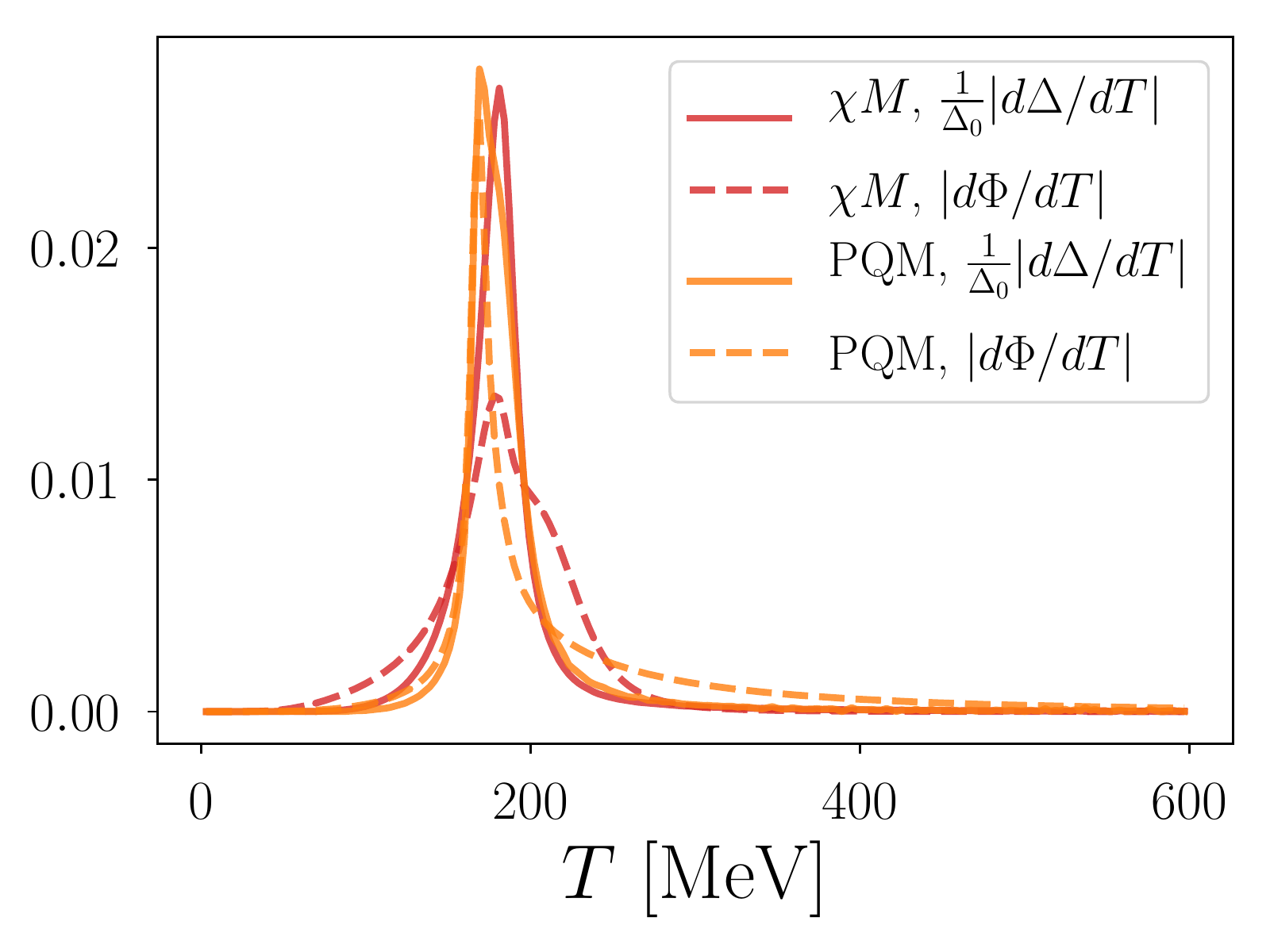}
\caption{Absolute value of the
  differentiated order parameters as 
function of $T$ for $\mu=0$. The peak locations correspond to the pseudocritical transition 
temperatures.}
		\label{fig:orderParams_b}
\end{figure}

One can define the pseudo-critical temperature
for example by the temperature at which the order parameter has
dropped to half its zero-temperature value. Another definition is
the temperature at which the derivative of the order parameter has
its peak. In this paper, we will stick to the latter.
In Fig.~\ref{fig:orderParams_b}, we show the derivatives of the
order parameters $\Delta(T)/\Delta(T=0)$ and $\Phi(T)$.
From the figure, we see that the pseudocritical temperatures 
for the chiral and deconfinement transitions coincide
for both models, with the inflection points of $\Delta$ being located at 
\bqa
T_{c}^{\chi M} = {181^{+6}_{-9}}\ \si{\mega\electronvolt}\;,
\hspace{0.6cm}
T_{c}^{\mathrm{PQM}} =  {169^{+3}_{-3}}\ \si{\mega\electronvolt}\;. \label{eq:modelTcs}
\eqa
The uncertainty is given by varying $\sigma$ from \SIrange{400}{550}{\mega\electronvolt}, with the lowest sigma mass corresponding to the lowest $T_c$ and vice versa.

\subsection{Pressure, energy density and interaction measure}
\begin{figure}[htb]
	\centering
\centering
\includegraphics[width=0.4\textwidth]{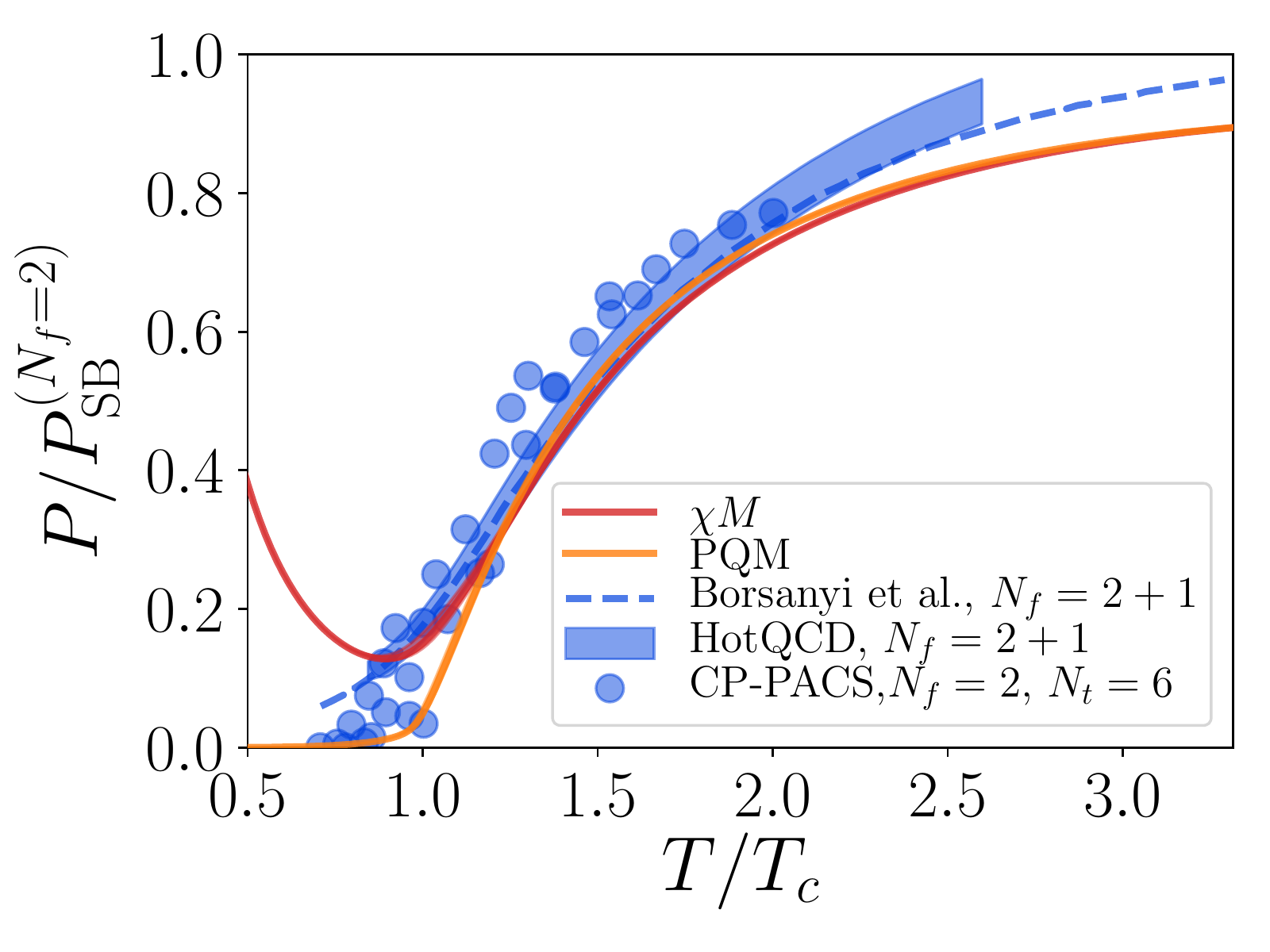}
	\caption{Boltzmann-normalized pressure as function of $T/T_c$ for $\mu=0$ in the \chiM{} and 
RRTW model compared to lattice data from 
\cite{HotQCD2014_thermodynamics, Wuppertal2014_thermodynamics, CPPACS_2flavor_thermodynamics}.}

		\label{fig:Pmu0}

\end{figure}

Let us now turn to the thermodynamic functions. 
Figure~\ref{fig:Pmu0} shows the comparison 
between the Stefan-Boltzmann (SB) normalized
pressure as calculated on the lattice \cite{HotQCD2014_thermodynamics, Wuppertal2014_thermodynamics, CPPACS_2flavor_thermodynamics} and in the two chiral models, with each data set plotted against $T/T_c$ for its respective $T_c$ (this also applies to plots in the following). For the $(2+1)$-flavor lattice data we normalize with $T_c = \SI{155}{\mega\electronvolt}$ \cite{Wuppertal2010_Tc, HotQCD2014_thermodynamics}, while for the model data the $T_c$s are given by \eqref{eq:modelTcs}. The two-flavor lattice data from Ref.~\cite{CPPACS_2flavor_thermodynamics} are obtained directly as function of $T/T_c$ without knowledge of $T_c$. The pressure in the model data and $(2+1)$-flavor lattice data is normalized with, 
\begin{align}
    P_{\mathrm{SB}}^{(N_f=2)} &= \left(\frac{2(N_c^2 - 1)}{90} + 2N_c N_f \frac{7}{360} \right)\pi^2 T^4 \nonumber \\ 
    &= \left(\frac{8}{45} + \frac{7}{30} \right)\pi^2 T^4 ,
\end{align}
while the two-flavor lattice data, which are not continuum extrapolated, are normalized with the  relevant Stefan-Boltzmann pressure for a discretized space-time (see Ref.~\cite{CPPACS_2flavor_thermodynamics} for details).
The uncertainty bands in the HotQCD data correspond to uncertainty in the continuum extrapolation. 
The uncertainty bands in the models are obtained by varying the sigma mass within the uncertainty 
range given in Ref~\cite{PDG}, which as mentioned is \SIrange{400}{550}{\mega\electronvolt}. The lowest $m_{\sigma}$ corresponds to the 
lowest temperature, and vice versa.

\begin{figure}[htb]
		\centering
		\includegraphics[width=0.4\textwidth]{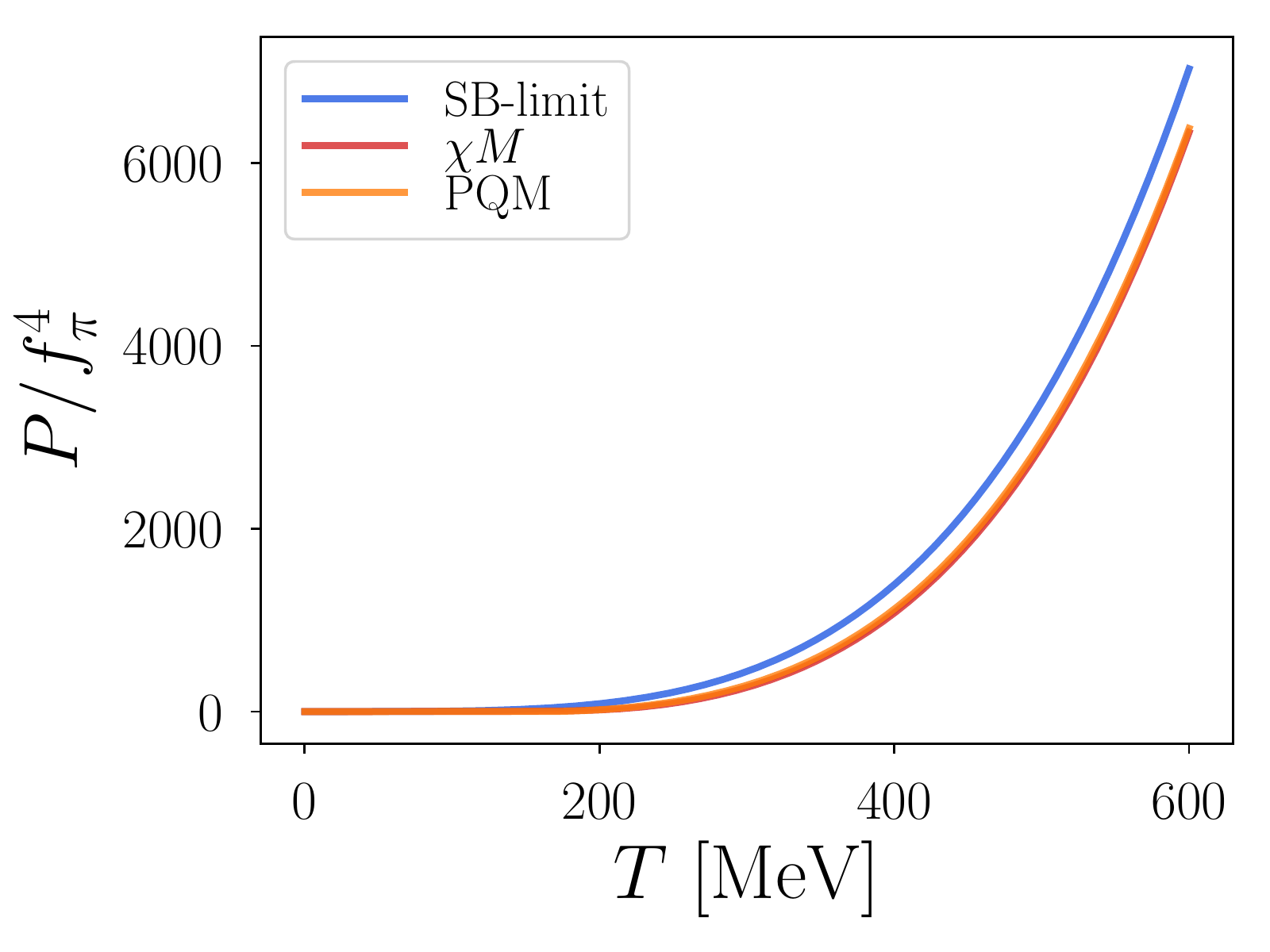}
	\caption{
Pressure normalized by $f_{\pi}^4$ at $\mu=0$
in the \chiM{} and PQM models compared to the SB-limit.}
		\label{fig:absPress}
\end{figure}

Both the PQM and \chiM{} models show reasonable agreement with lattice data 
above $T=T_c$, although with a slightly lower pressure. Below and 
around $T=T_c$ the PQM model appears to have a pressure that is significantly 
lower than what lattice data show. However, below and around $T_c$ we expect 
mesons to exist and contribute to the pressure, and by neglecting mesonic 
fluctuations in the model, we have underestimated the pressure. 
Since the pions have masses of $\sim \SI{140}{\mega\electronvolt}$ below $T_c$ 
while the quarks have masses of $\sim \SI{300}{\mega\electronvolt}$, we expect 
that the mesons would provide a significant contribution to the pressure in 
this range. For temperatures below $T_c$, the agreement with the \chiM{} model 
is worse, since there is a small but nonzero pressure causing $P/P_{\mathrm{SB}}$ 
to blow up for low temperatures due to the $T^4$-dependence of $P_{\rm SB}$.
However, this does not mean that the pressure diverges or that it is
large. It only means that a small non-zero pressure exists for $T>0$. This 
pressure is insignificant, as we see if we compare the pressure of the \chiM{} 
and the RRTW models without SB-normalizing, which is done in 
Fig.~\ref{fig:absPress}. 

\begin{figure}[htb]
	\centering
		\centering
		\includegraphics[width=0.4\textwidth]{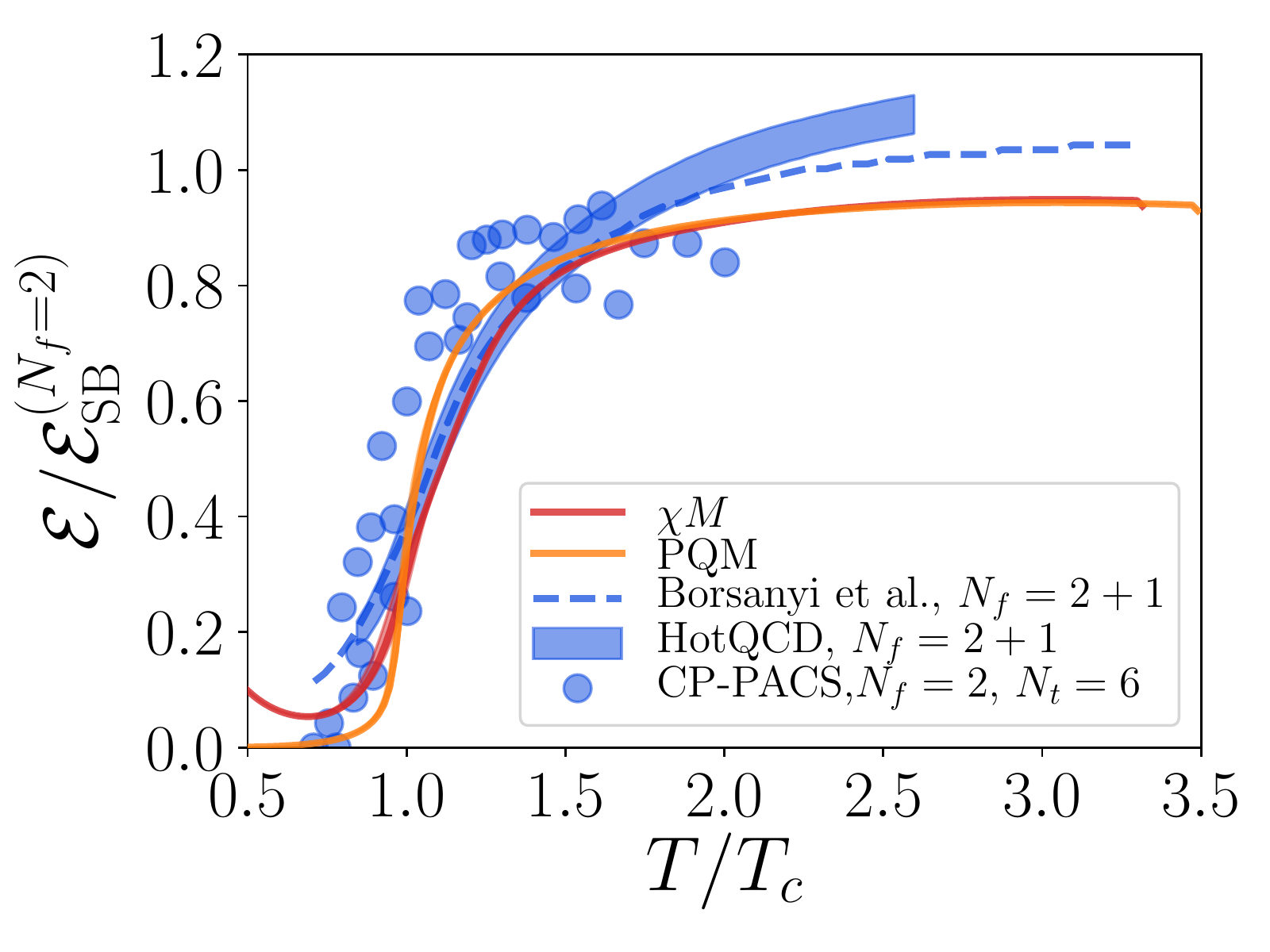}
		\centering
		\includegraphics[width=0.4\textwidth]{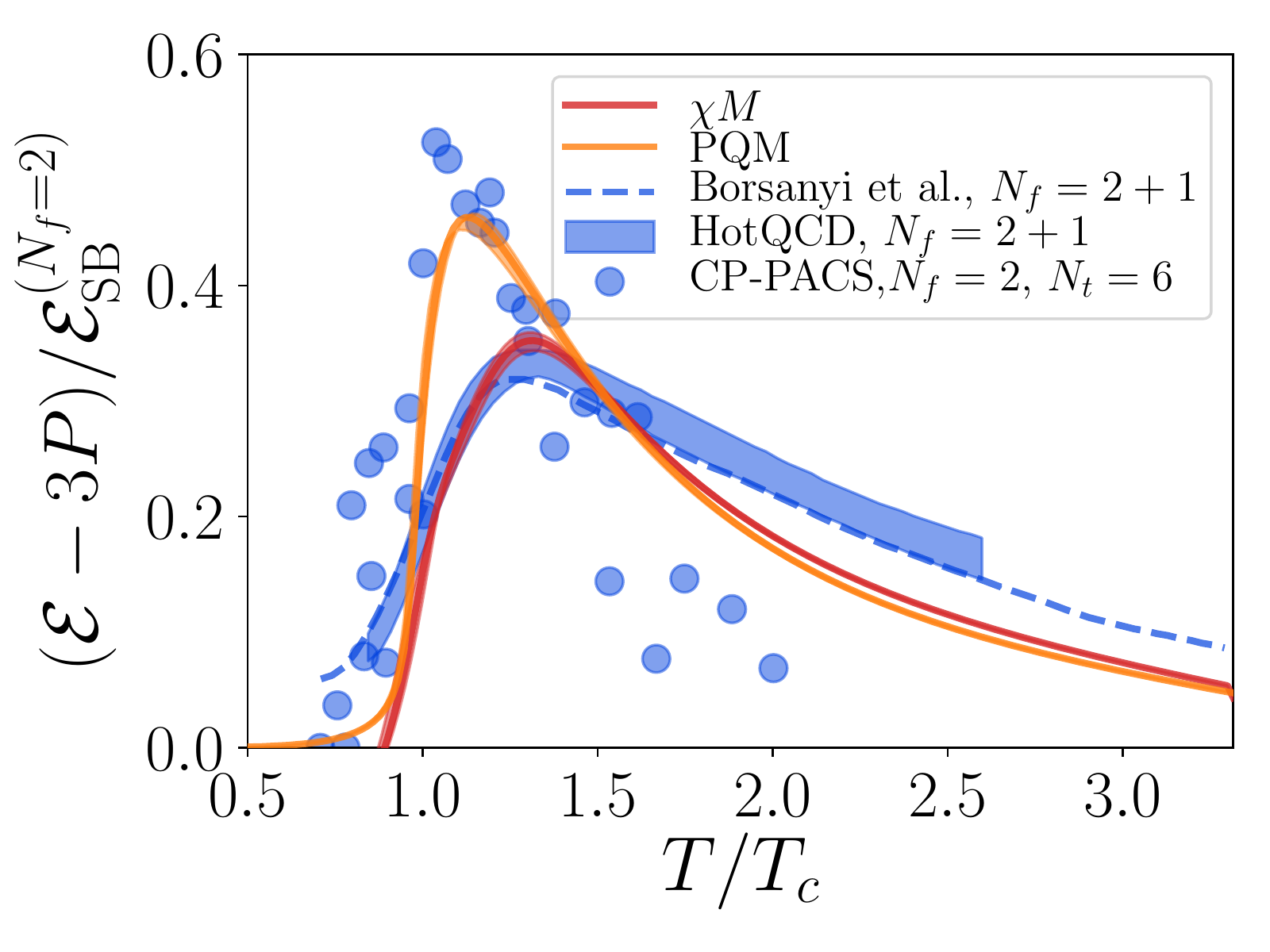}
	\caption{SB-normalized energy density (upper panel) and interaction measure (lower panel) 
	at $\mu=0$ in the effective models compared to lattice data from \cite{HotQCD2014_thermodynamics, Wuppertal2014_thermodynamics, CPPACS_2flavor_thermodynamics}.}
	\label{fig:edensAndAnomaly}
\end{figure}

The energy density $\mathcal{E}$ and the interaction measure $I$, both 
normalized with $\mathcal{E}_{\mathrm{SB}}^{(N_f = 2)}=3P_{\mathrm{SB}}^{(N_f = 2)}$, are
shown in Fig. \ref{fig:edensAndAnomaly} 
(with error bars are obtained in the same way as for the pressure). We find 
fairly good agreement between the PQM model and two-flavor lattice data up to 
$T\sim 1.5 T_c$.
The peak of the interaction measure in the \chiM{} model is shifted to higher 
values than what is seen in the PQM model and two-flavor lattice data. 
The \chiM{} model also has an interaction measure that is negative for low 
temperatures and a peak that is too low. 

\begin{figure}[htb]
	\centering
		\includegraphics[width=0.4\textwidth]{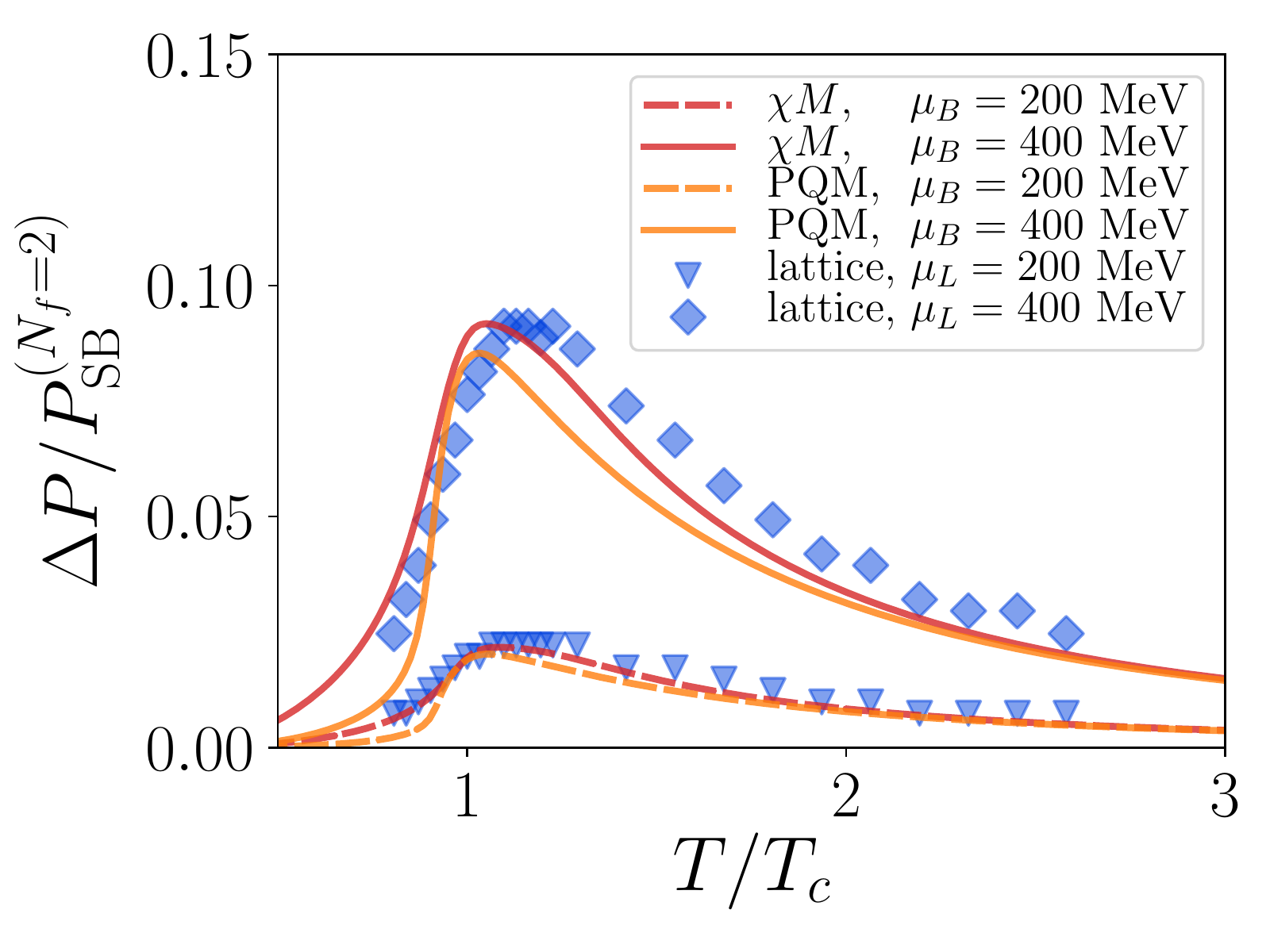}
		\includegraphics[width=0.4\textwidth]{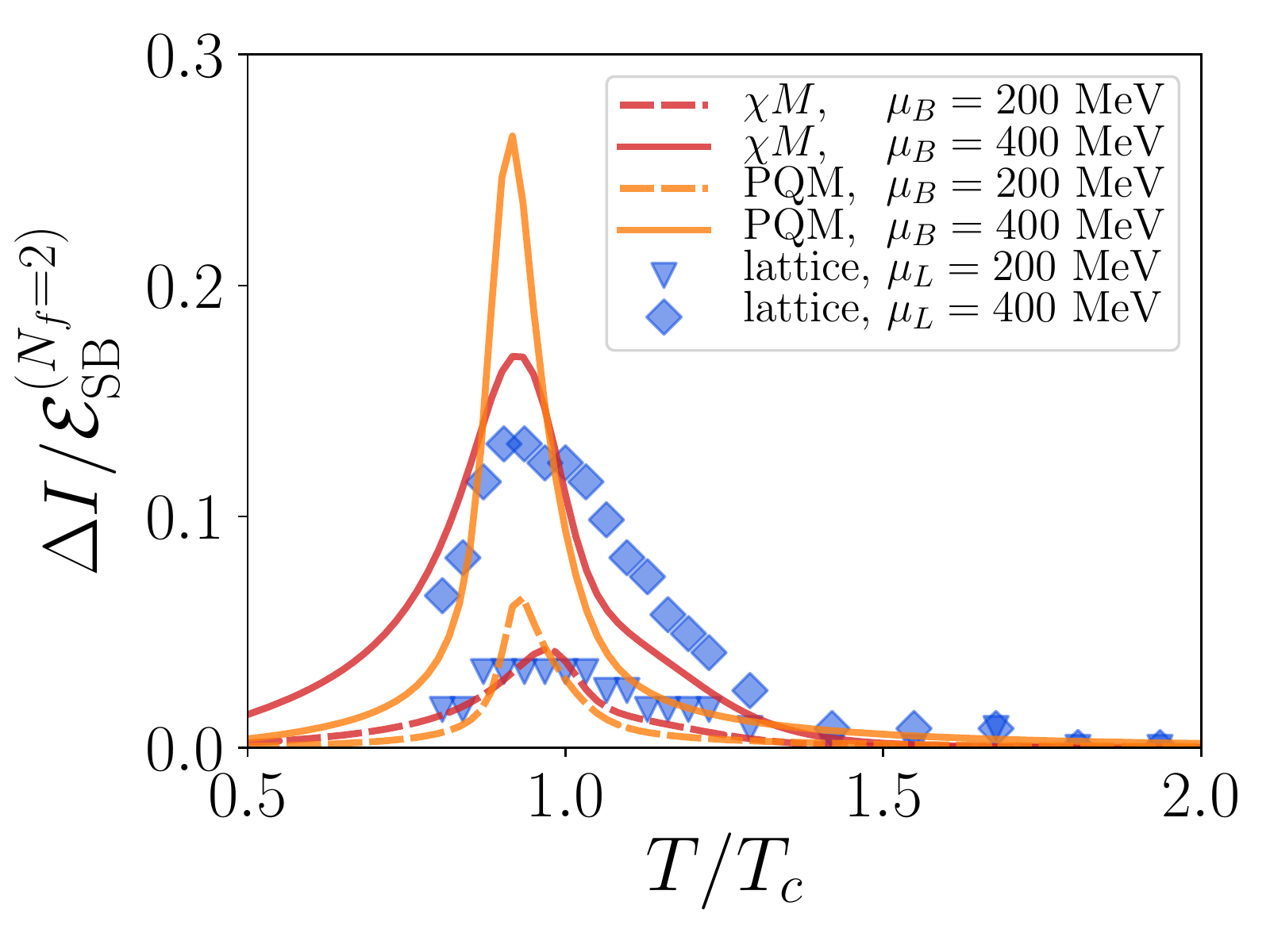}
	\caption{SB-normalized increase in pressure (upper panel) and interaction measure (lower panel) in the effective models at nonzero baryon chemical potential compared to lattice data from \cite{Wuppertal_finiteMu}. See main text for details.}
	\label{fig:press_anomaly_mu}
\end{figure}

In Fig.~\ref{fig:press_anomaly_mu} we plot the Boltzmann-normalized version of the quantities
\begin{align}
\Delta P(\mu, T) &= P(\mu, T) - P(0, T)\;, \\
\Delta I(\mu, T) &= I(\mu, T) - I(0, T)\;,
\end{align}
at $\mu_B = 3\mu = \SI{200}{\mega\electronvolt}$ and $\mu_B =\SI{400}{\mega\electronvolt}$. The model data are compared with $(2+1)$-flavor lattice data from Ref.~\cite{Wuppertal_finiteMu}. We compare with lattice data where the chemical potentials for the light flavors are $\mu_L = 3\mu_u = 3\mu_d = \SI{200}{\mega\electronvolt}$ and $\mu_L = \SI{400}{\mega\electronvolt}$, while the strange chemical potential is chosen so that the net strangeness density is zero. Since $\mu_s \neq \mu_d = \mu_u$, we use the notation from Ref.~\cite{Wuppertal_finiteMu} and denote $3\mu_d = 3\mu_q$ as $\mu_L$ instead of $\mu_B$ in the $(2+1)$-flavor simulation. We compare with lattice data where net strangeness density is zero, since this scenario should resemble the two-flavor situation more than when $\mu_u = \mu_d = \mu_s$, for which the strangeness is nonzero.

We see that the pressure, which increases as function of temperature at nonzero baryon chemical potential, agrees fairly well with lattice data for both models. For the interaction measure, we find that the PQM model has a significantly higher peak than lattice data at $\mu_B = \SI{400}{\mega\electronvolt}$, while the \chiM{} model is in better agreement. 

\subsection{Phase diagram}
In this subsection we present various phase diagrams of the two
models. We first discuss the different phases in the $\mu$-$T$ plane.
Then we move on to the phase diagram in the $\mu_I$-$T$ plane, where
we include the possibility of condensation of charged pions.
In calculating the phase diagrams, we have dropped the $\mathcal{U}_{q, \mathrm{cur}}$ term, since its effect on thermodynamics and critical temperatures is found to be entirely negligible.

\begin{figure}[htb]
	\centering
		\centering
	\includegraphics[width=0.5\textwidth]{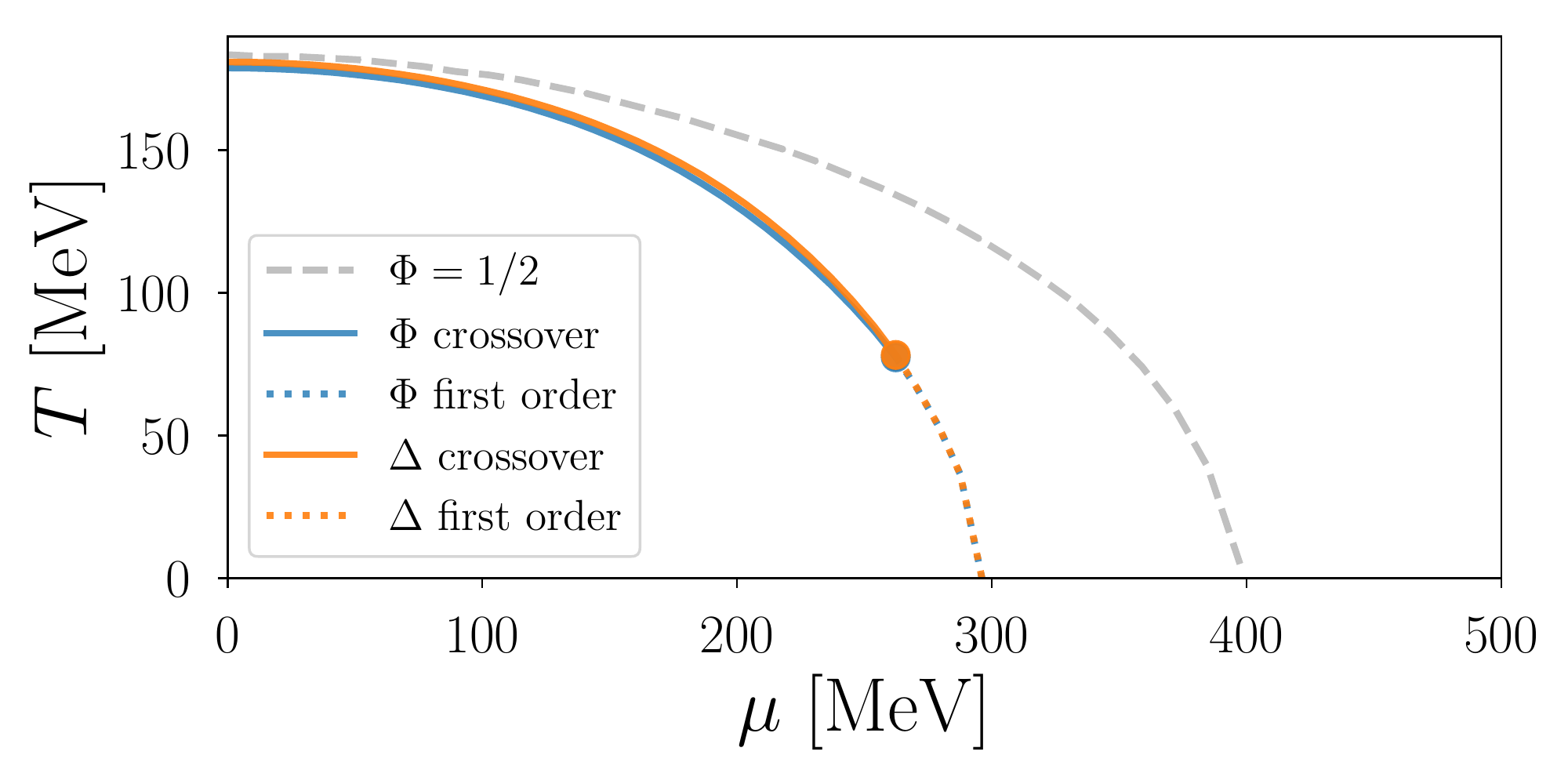}
		\caption{Phase diagram for the \chiM{} model in the
$\mu$--$T$ plane.}
\label{half1}

\end{figure}

\begin{figure}[htb]
		\centering
	\includegraphics[width=0.5\textwidth]{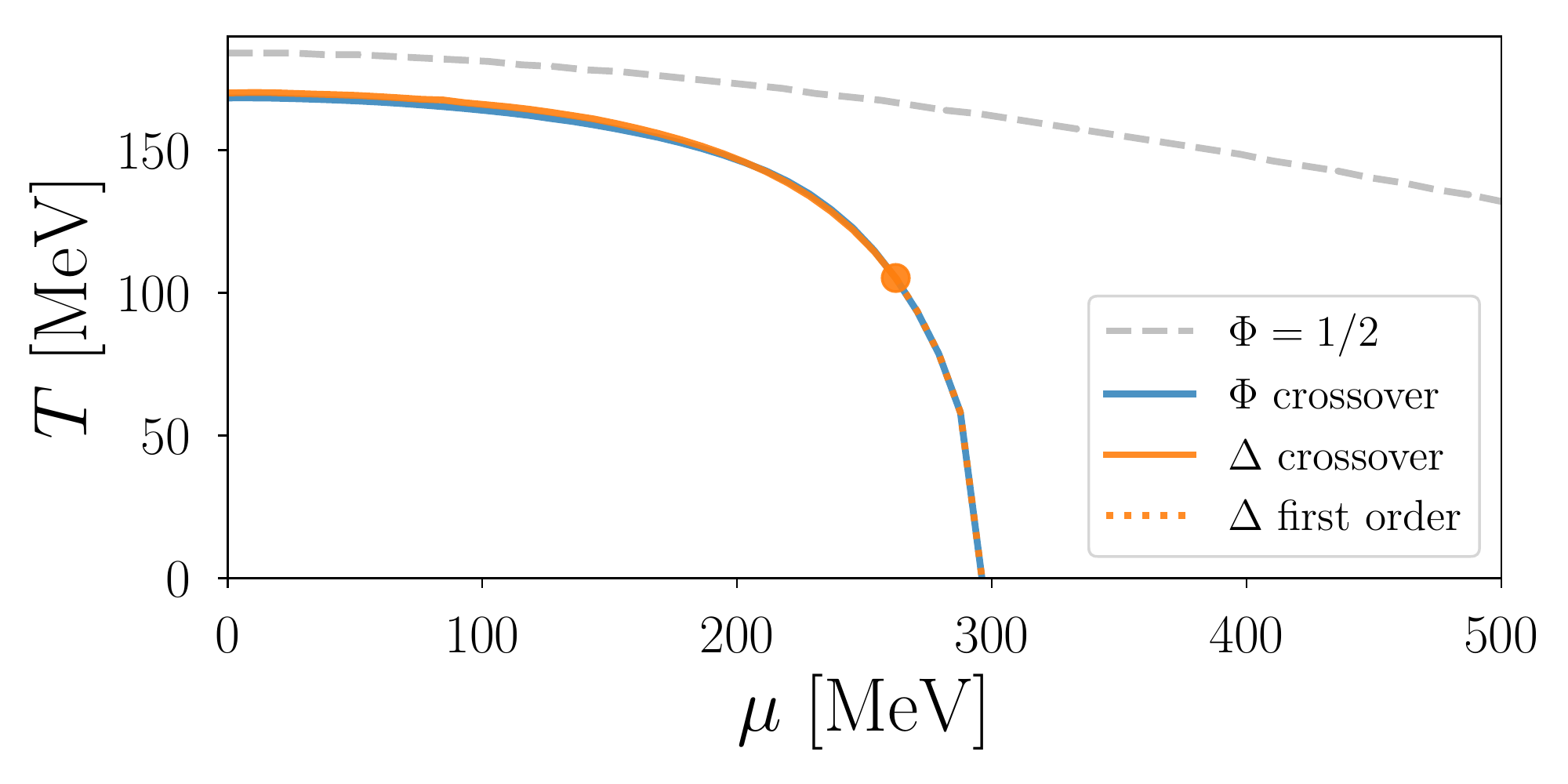}
		\caption{Phase diagram for the PQM model in the
$\mu$--$T$ plane.}
\label{half2}
\end{figure}

Figures \ref{half1}
and \ref{half2} display the phase diagrams for the two models, where the 
pseudocritical temperatures corresponding to the inflection points of $\Delta$ 
and $\Phi$ 
are indicated, in addition to the temperature where $\Phi ={1\over2}$. 


\begin{figure*}[htb!]
    \includegraphics[width=0.9\textwidth]{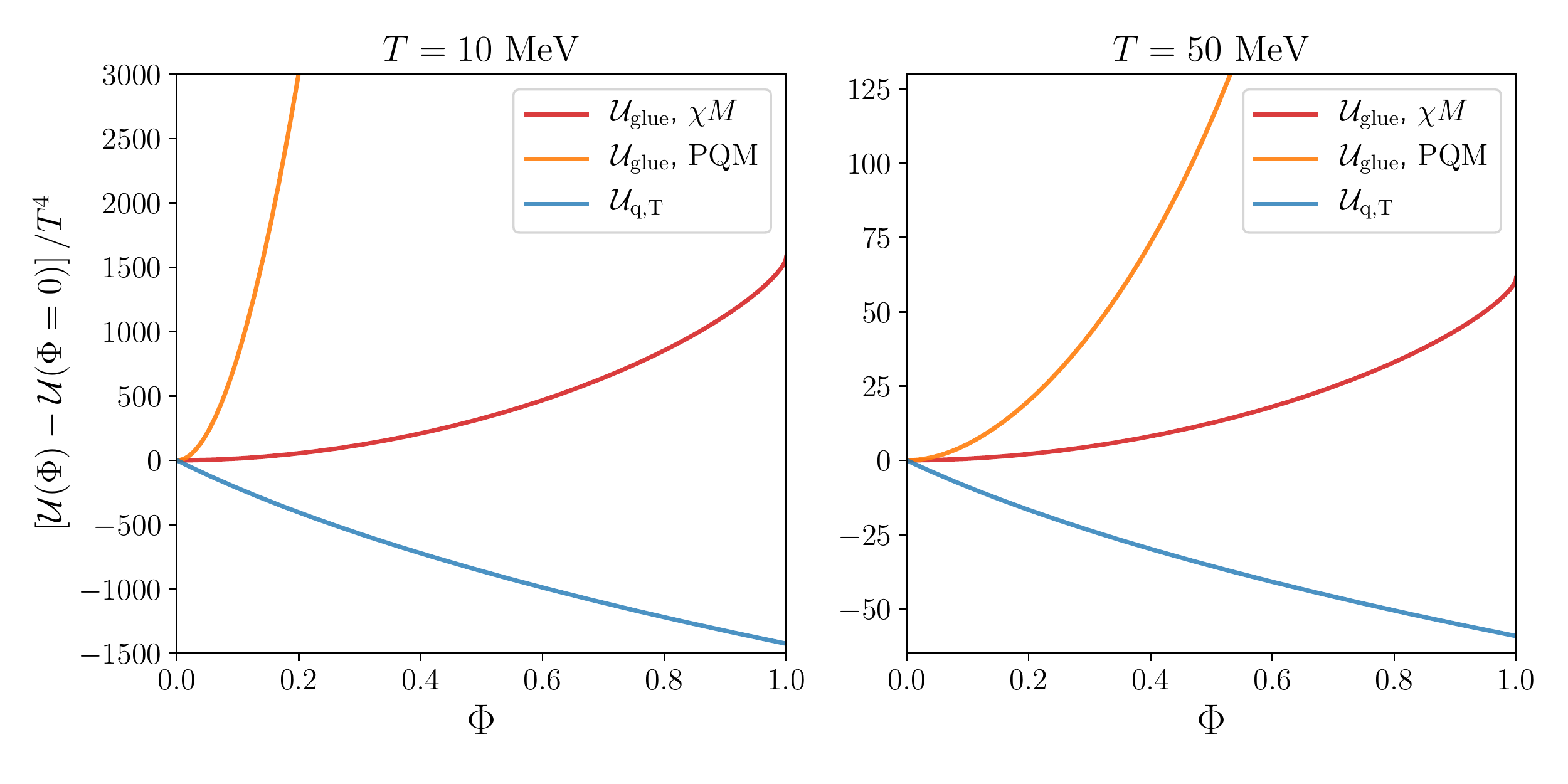}
    \caption{The gluonic and quark potentials as function of the Polyakov loop at $\mu = \SI{400}{\mega\electronvolt}$, $\Delta = \SI{400}{\mega\electronvolt}$ and $r=0$. }
    \label{fig:gluonPot}
\end{figure*}

\begin{figure}[htb!]
	\centering
		\centering
		\includegraphics[width=0.4\textwidth, trim={.65cm, 0.7cm, 2.2cm, 2.1cm}, clip=true]{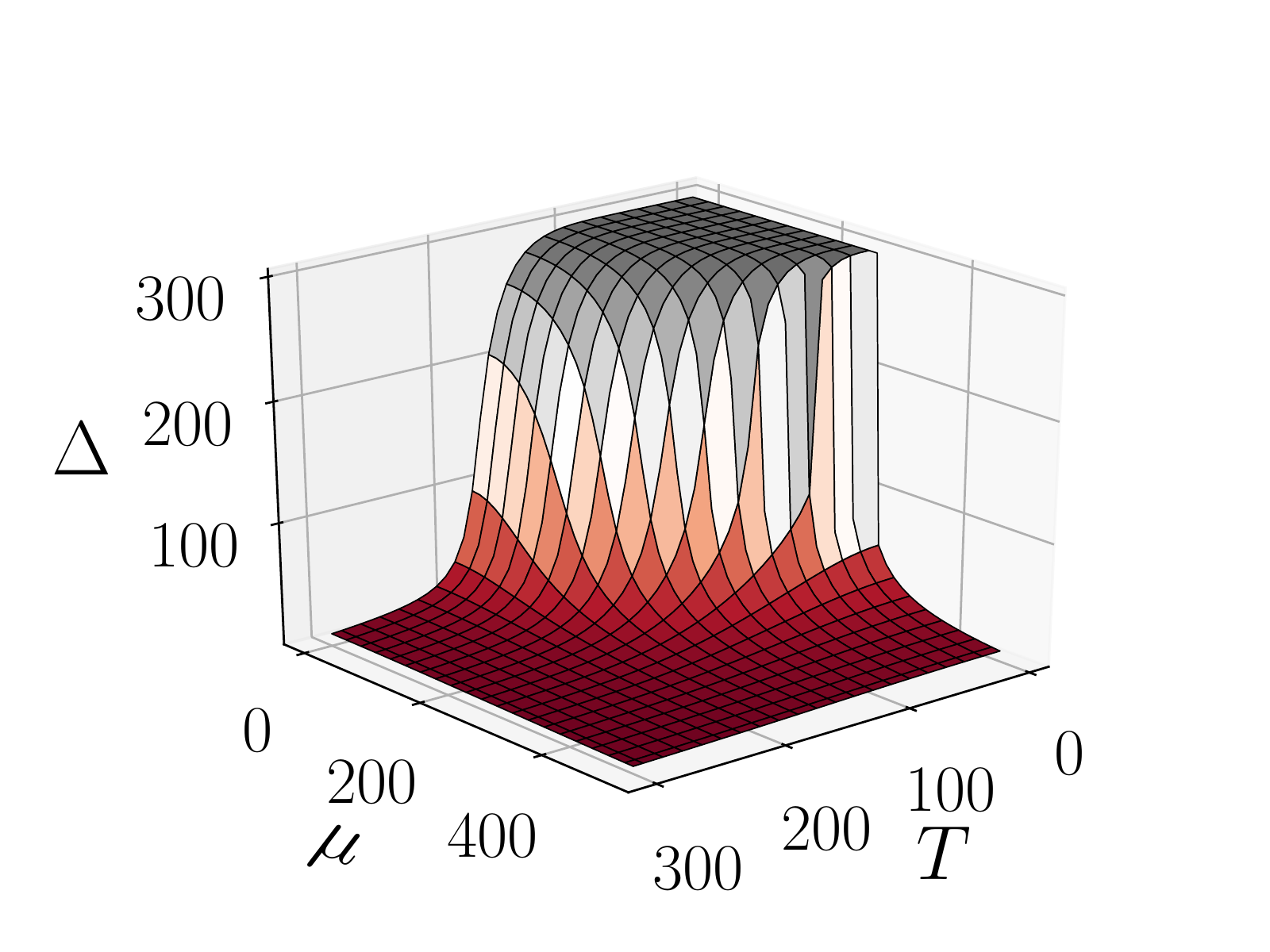}
	\caption{The chiral condensate as function of $(\mu, T)$ in the chiral
matrix model. Units of $\Delta$, $\mu$ and $T$ are $\mathrm{MeV}$.}
	\label{fig:3D_deltachi}
\end{figure}

\begin{figure}[htb!]
		\centering
		\includegraphics[width=0.4\textwidth, trim={.65cm, 0.7cm, 2.2cm, 2.1cm}, clip=true]{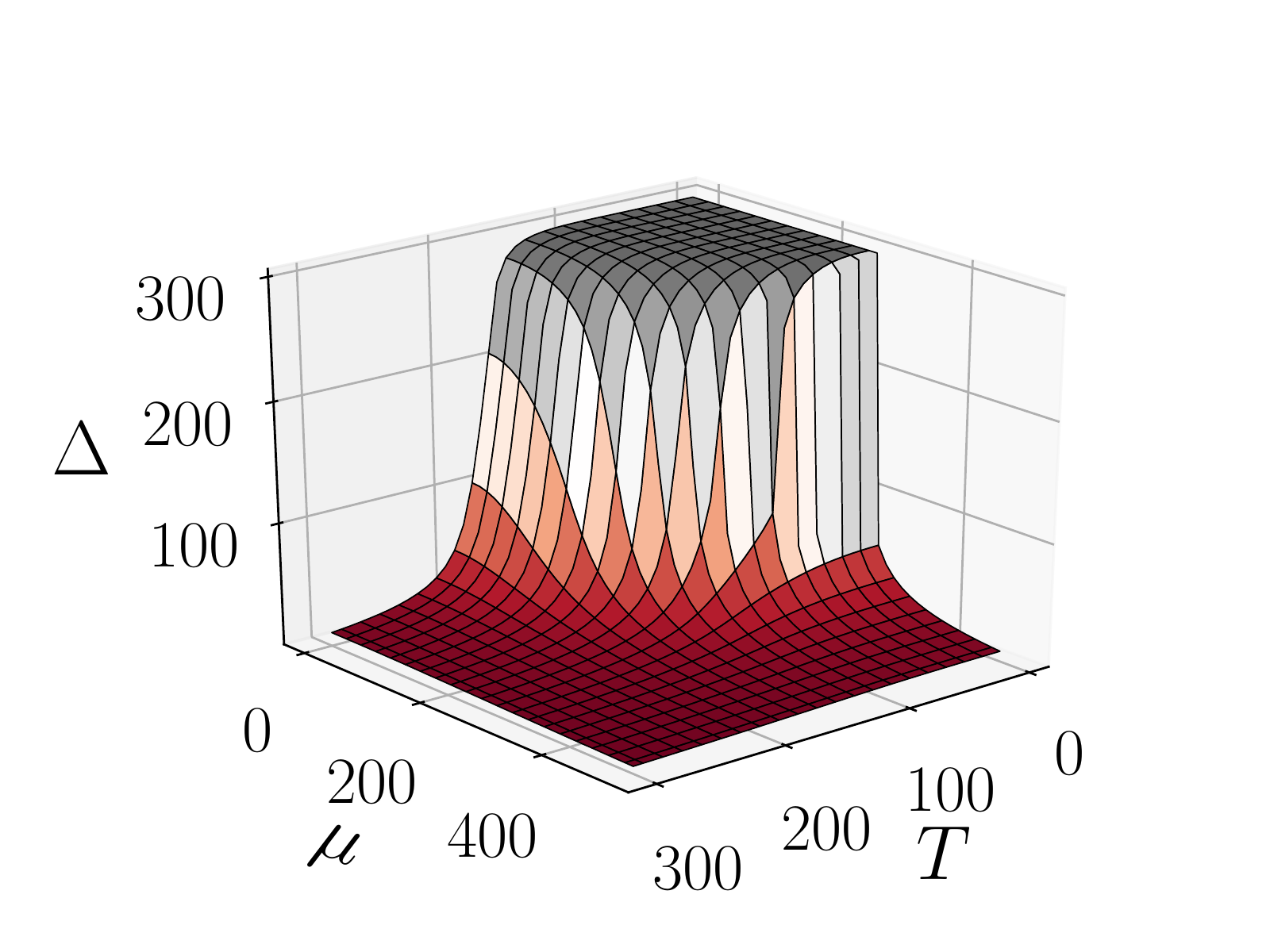}
	\caption{The chiral condensate as function of $(\mu, T)$ in the PQM
model. Units of $\Delta$, $\mu$ and $T$ are $\mathrm{MeV}$.}
	\label{fig:3D_deltapqm}
\end{figure}

\begin{figure}[htb]
	\centering
		\includegraphics[width=0.40\textwidth2, trim={.65cm, 0.7cm, 2.0cm, 2.1cm}, clip=true]{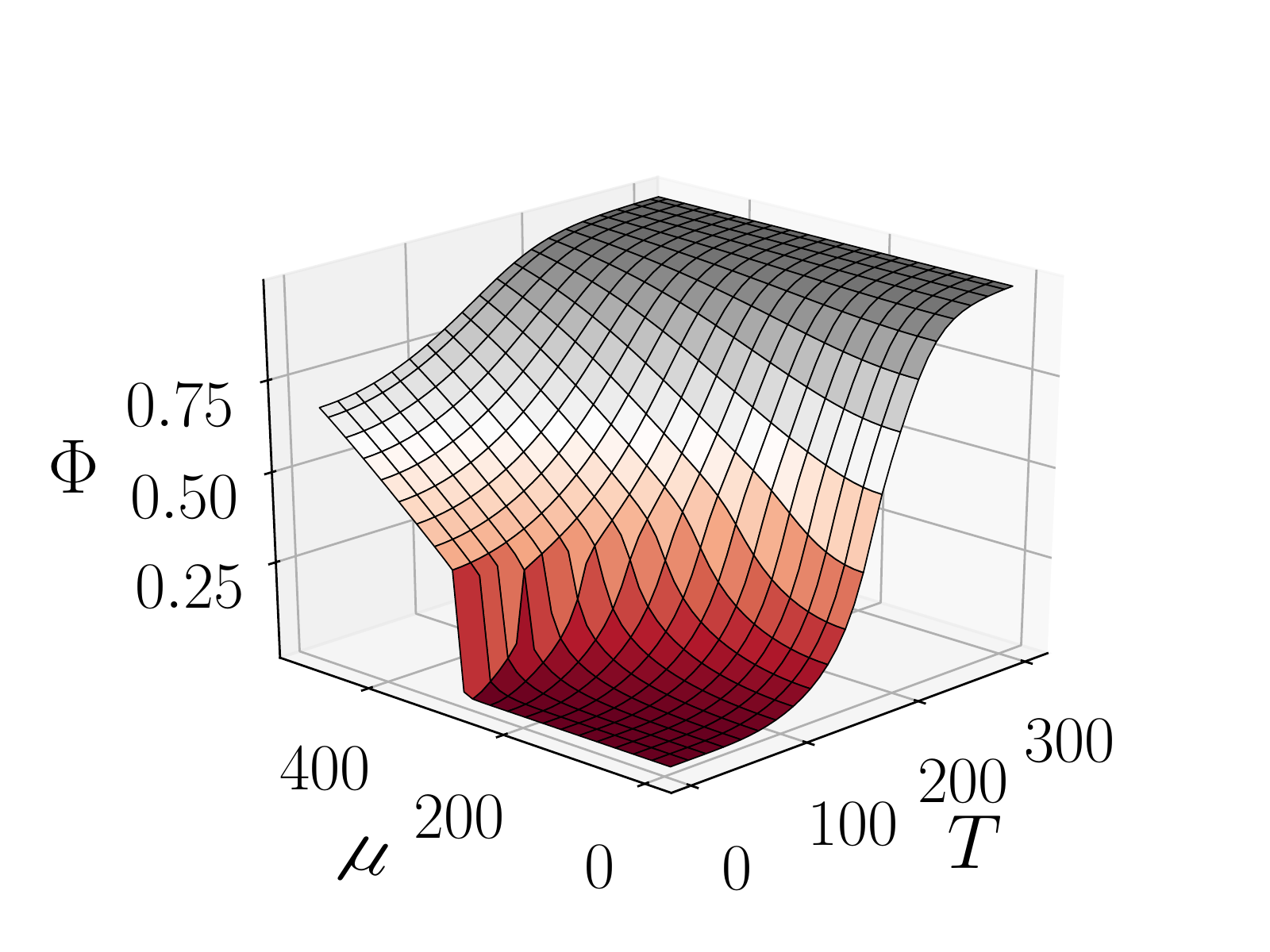}
	\caption{The Polyakov loop as function of $(\mu, T)$ in the chiral
matrix model. 
Units of $\mu$ and $T$ are $\mathrm{MeV}$.}
	\label{fig:3D_polyakovchi}
\end{figure}

\begin{figure}[htb]
		\includegraphics[width=0.40\textwidth, trim={.65cm, 0.7cm, 2.0cm, 2.1cm}, clip=true]{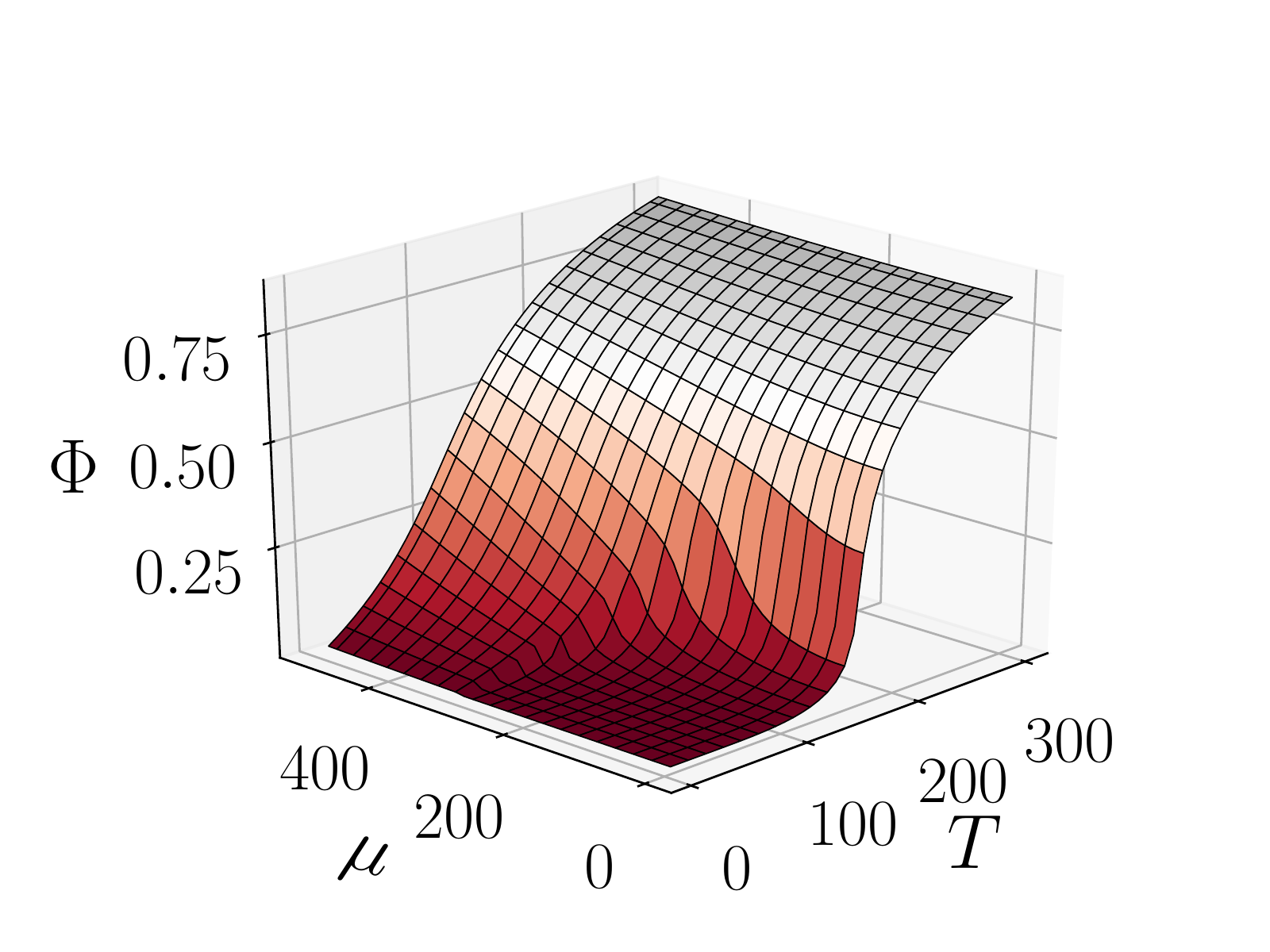}
	\caption{The Polyakov loop as function of $(\mu, T)$ in the PQM model. 
Units of $\mu$ and $T$ are $\mathrm{MeV}$.}
	\label{fig:3D_polyakovpqm}
\end{figure}

We see that the chiral and deconfinement phase transitions happen roughly  simultaneously also for nonzero chemical potentials. 
Note however that referring to the inflection point of the Polyakov loop as 
``deconfinement'' in the regime of high chemical potential is misleading. It is 
correct that the chiral symmetry in the models is approximately restored 
above the orange lines in Figs.~\ref{half1} and \ref{half2}, since we can see from 
Figs.~\ref{fig:3D_deltachi} and \ref{fig:3D_deltapqm} that $\Delta \rightarrow 0 $ quickly for temperatures higher than the crossover temperature $T_c$. 
However, it is not correct to assume that quarks are 
deconfined everywhere outside the phase boundaries, since the inflection point of the Polyakov loop can be relatively far away from the region where center symmetry is approximately restored ($\Phi\approx 1$). This is visible from 
Figs.~\ref{fig:3D_polyakovchi} and~\ref{fig:3D_polyakovpqm}, which show the 
value of the Polyakov loop in the 
$\mu-T$ plane. We see in the PQM model that the Polyakov loop is close to the 
confining value of $\Phi=0$ also for $\mu > {300}{\rm MeV}$, given
low temperatures. Interestingly, we see that we approach deconfinement in the 
\chiM{} model in the high-density limit, which is not the case in the PQM
model. This is a major difference between the two models. 
The difference stems from the fact that at low temperatures the value of the gluon potential as function of $\Phi$ away from its minimum grows significantly faster in the PQM model than in the \chiM{} model, so at low temperatures the gluonic potential strongly dominates in the PQM model. The free energy gained from the quark potential by deconfining is negligible compared to the gluon energy cost. This is however not the case of the \chiM{} model. This is clear from Fig.~\ref{fig:gluonPot}, where we see that the \chiM{} gluon potential is much flatter around its minimum than the PQM model. The flatness of the gluonic potential of the \chiM{} model causes the deconfinement transition to track the chiral transition to a larger degree.
It is hard to assess which behavior best reflects QCD due to the lack of lattice data in that region. 

We also note that at sufficiently large temperatures, some of the crossovers 
become first order phase transitions, with the transition from crossover to 
first order marked by a critical point. In the \chiM{} 
model the critical points of 
of the two transition lines coincide, while for the PQM 
model only the line of chiral transition has a critical point. This is another 
qualitative difference between the two models. 


\subsection{Pion condensation}
We now move on to discuss 
the phase diagram in the $\mu_I$-$T$ plane, which requires a treatment of Bose condensation of charged pions.
For simplicity, we set the baryon chemical potential to zero in the
remainder of this section. 

In addition to an expectation value of $\tilde{\sigma}$ we now allow for a nonzero
expectation value of $\sqrt{\pi_1^2 + \pi_2^2}$ denoted by $\pi_0$. Introducing
$\rho=g\pi_0$ in analogy with $\Delta$, the tree-level
potential can be written as
\begin{equation}
    \mathcal{U}_{\rm tree} = 
    {1\over2}{m^2\over g^2}\Delta^2+{1\over2}{m^2-4\mu_I^2\over g^2}\rho^2
    +{\lambda\over24g^4}\Delta^4-{h\over g}\Delta\;.
\end{equation}
The quark energies can be read off from the zeros of the quark determinant and read
\bqa
\label{eu}
E_u&=&E(-\mu_I)\;,
\hspace{1cm}
E_d=E(\mu_I)\;,
\\
E_{\bar{u}}&=&E(\mu_I)\;,
\hspace{1.25cm}E_{\bar{d}}=E(-\mu_I)\;.
\label{ed}
\eqa
where we have defined
\bqa
E(\mu_I)&=&
\left[\left(\sqrt{\pp^2+\Delta^2}+{\mu_I}\right)^2+\rho^2\right]^{1\over2}\;.
\eqa
The effective potential at $T=0$ then is
\bqa
\mathcal{U}_{\mathrm{vac}}&=&\mathcal{U}_{\mathrm{tree}} -N_c\int_p\left(E_{u}+E_{d}+E_{\bar{u}}+E_{\bar{d}}\right)\;,
\label{emu}
\eqa
where the last term is the one-loop contribution. 
It cannot be evaluated analytically for nonzero $\rho$.
In Ref. \cite{patrick}, it was evaluated by isolating
the divergent pieces and writing $\mathcal{U}_{\mathrm{vac}}=V_{\rm div}+V_{\rm fin}$. 
The divergent term $V_{\rm div}$
was then evaluated using dimensional regularization, and the poles in
$\epsilon$ (evaluating integrals in $d=3-2\epsilon$ dimensions) were
removed by renormalization
using the $\overline{\rm MS}$ scheme in the usual way. The running
parameters were finally eliminated in favor of the physical masses
and the pion-decay constant. The final result is
\begin{widetext}
\bqa\nonumber
\mathcal{U}_{\rm vac}&=&
\dfrac{3}{4}m_\pi^2 f_\pi^2
\left\{1-\dfrac{4 m_q^2N_c}{(4\pi)^2f_\pi^2}m_\pi^2F^{\prime}(m_\pi^2)
\right\}\dfrac{\Delta^2+\rho^2}{m_q^2}
\\ \nonumber &&
 -\dfrac{1}{4}m_\sigma^2 f_\pi^2
\left\{
1 +\dfrac{4 m_q^2N_c}{(4\pi)^2f_\pi^2}
\left[ \left(1-\mbox{$4m_q^2\over m_\sigma^2$}
\right)F(m_\sigma^2)
 +\dfrac{4m_q^2}{m_\sigma^2}
-F(m_\pi^2)-m_\pi^2F^{\prime}(m_\pi^2)
\right]\right\}\dfrac{\Delta^2+\rho^2}{m_q^2} 
\\ \nonumber &&
-2\mu_I^2f_\pi^2
\left\{1-\dfrac{4 m_q^2N_c}{(4\pi)^2f_\pi^2}
\left[\log\mbox{$\Delta^2+\rho^2\over m_q^2$}
+F(m_\pi^2)+m_\pi^2F^{\prime}(m_\pi^2)\right]
\right\}{\rho^2\over m_q^2}
\\ \nonumber
 & & + \dfrac{1}{8}m_\sigma^2 f_\pi^2
\left\{ 1 -\dfrac{4 m_q^2  N_c}{(4\pi)^2f_\pi^2}\left[
\dfrac{4m_q^2}{m_\sigma^2}
\left( 
\log\mbox{$\Delta^2+\rho^2\over m_q^2$}
-\mbox{$3\over2$}
\right) -\left( 1 -\mbox{$4m_q^2\over m_\sigma^2$}\right)F(m_\sigma^2)
+F(m_\pi^2)+m_\pi^2F^{\prime}(m_\pi^2)\right]
 \right\}\dfrac{(\Delta^2+\rho^2)^2}{m_q^4}
\\ &&
- \dfrac{1}{8}m_\pi^2 f_\pi^2
\left[1-\dfrac{4 m_q^2N_c}{(4\pi)^2f_\pi^2}m_\pi^2F^{\prime}(m_\pi^2)\right]
\dfrac{(\Delta^2+\rho^2)^2}{m_q^4}
-m_\pi^2f_\pi^2\left[
1-\dfrac{4 m_q^2  N_c}{(4\pi)^2f_\pi^2}m_\pi^2F^{\prime}(m_\pi^2)
\right]\dfrac{\Delta}{m_q}
+V_{\rm fin}\;,
\label{fullb}
\eqa
where the finite contribution $V_{\rm fin}$ is 
\bqa
V_{\rm fin}&=&
-N_c\int_p\left(E_{u}+E_{d}+E_{\bar{u}}+E_{\bar{d}}\right)+
4N_c\int_p\left[
\sqrt{p^2+\Delta^2+\rho^2}
+{\mu_I^2\rho^2\over2({p^2+\Delta^2+\rho^2)^{3\over2}}}\right]\;,
\eqa
which must be evaluated numerically. Eq.~(\ref{fullb}) reduces
to Eq.~(\ref{eq:renormGrandPot}) for $\rho=0$; this can be easily seen
by noting that $V_{\rm fin}=0$ in this case.

The medium-dependent part of the one-loop effective potential at $\mu_B=0$ is
\bqa\nonumber
\mathcal{U}_{q, T}&=&-2T\int{d^3p\over(2\pi)^3}
\bigg\{
\log\Big[1+3(\Phi+\bar{\Phi}e^{-\beta E_u})e^{-\beta E_u}+e^{-3\beta E_u}\Big]
+\log\Big[1+3(\bar{\Phi}+\Phi e^{-\beta E_{\bar{u}}})e^{-\beta E_{\bar{u}}}
+e^{-3\beta E_{\bar{u}}}\Big]
\\
&&
+\log\Big[1+3(\Phi+\bar{\Phi}e^{-\beta E_d})e^{-\beta E_d}+e^{-3\beta E_d}\Big]
+\log\Big[1+3(\bar{\Phi}+{\Phi} e^{-\beta E_{\bar{d}}})e^{-\beta E_{\bar{d}}}
+e^{-3\beta E_{\bar{d}}}\Big]
\bigg\}\;.
\label{real}
\eqa
\end{widetext}
Note that this term vanishes at $T=0$.
As discussed previously, we see that two and two terms are complex conjugates of each other, and $\mathcal{U}_{q, T}$ is thus real, reflecting that there is
no sign problem when $\mu_B=0$.

In Ref. \cite{patrick}, it was shown that the zero-temperature effective
potential (\ref{fullb}) exhibits a second-order phase transition at
exactly $\mu_I^c={1\over2}m_{\pi}$. This was done by expanding $\mathcal{U}_{\rm vac}$
in powers of $\rho$, $\mathcal{U}_{\rm vac}=\alpha_0+\alpha_2\rho^2+\alpha_4\rho^4$
evaluating it at $\Delta=m_q$ (i.e.\ in the vacuum). The critical
chemical potential is defined by $\alpha_2=0$. The transition is second
order at $\mu_I=\mu_I^c$ since $\alpha_4$ was found to be positive
for this value of the isospin chemical potential. 

\begin{figure}[htb]
  \includegraphics[width=0.4\textwidth]{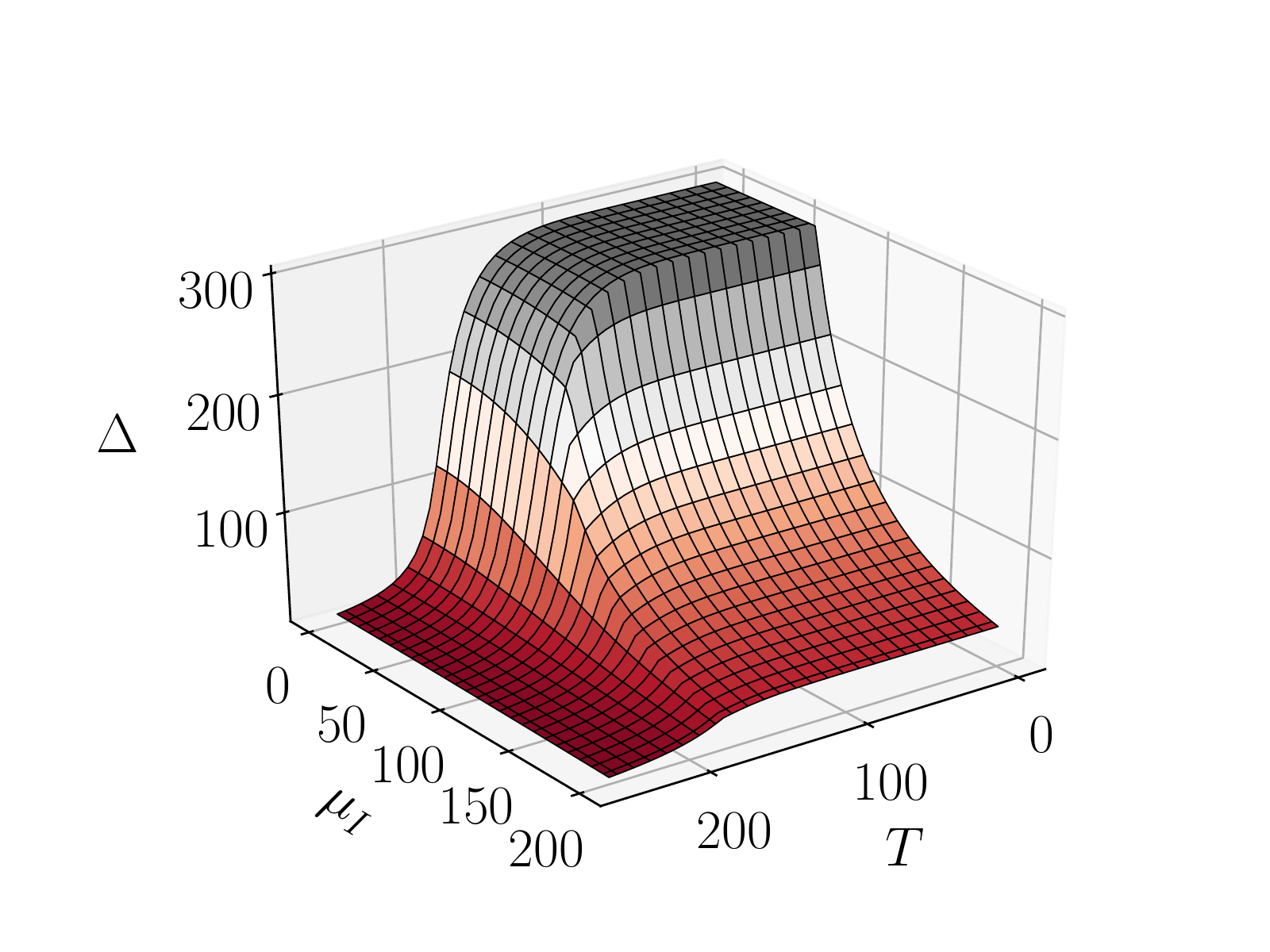}
\caption{Chiral condensate $\Delta$ as function of $\mu_I$ and $T$ for the \chiM{} model. Units of $\Delta$, $\mu_I$ and $T$ are $\mathrm{MeV}$.}
\label{fig:isoDelta3D}
\end{figure}

\begin{figure}[htb]
  \includegraphics[width=0.4\textwidth]{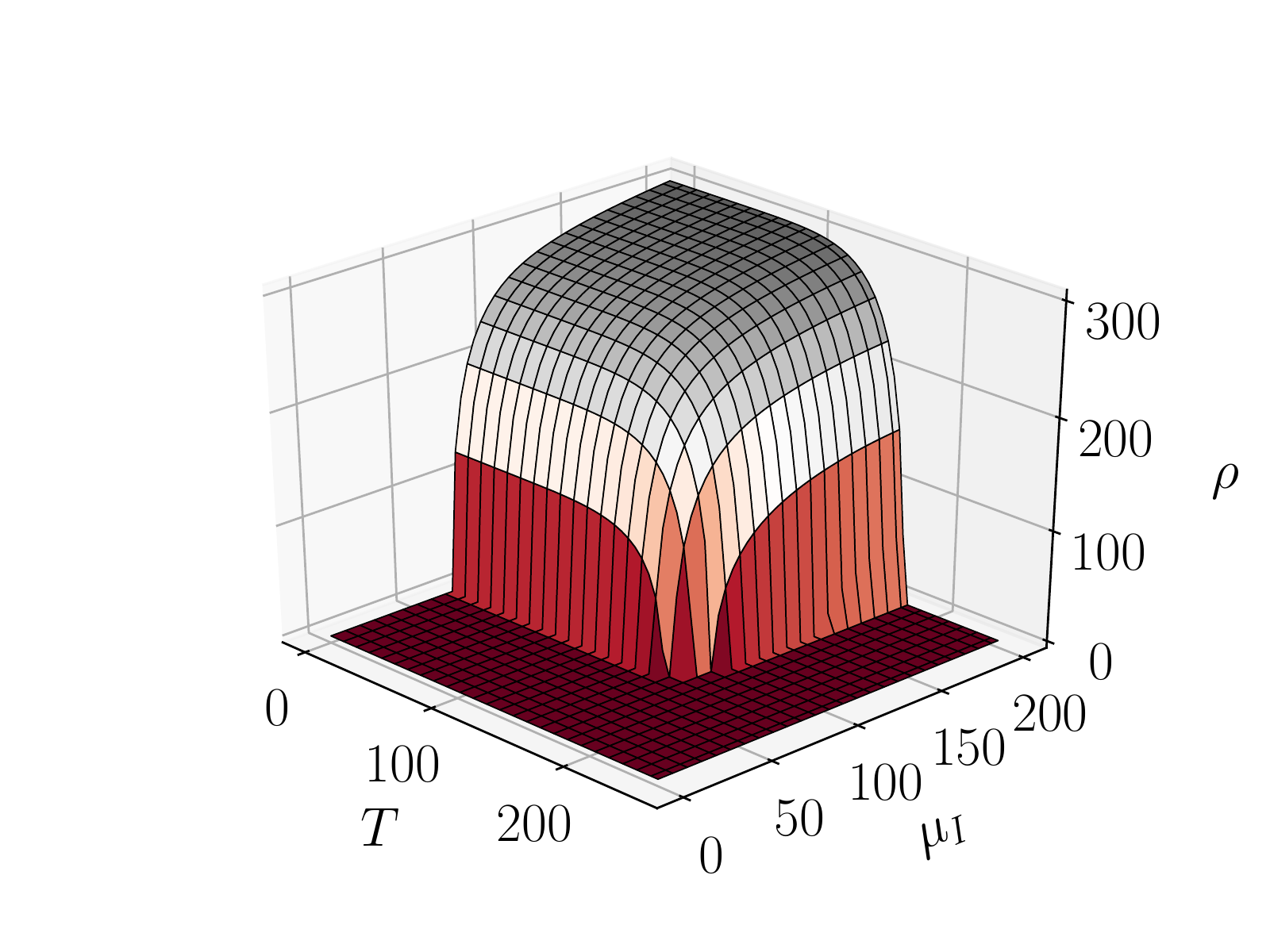}
\caption{Pion condensate $\rho$ as function of $\mu_I$ and $T$ for the \chiM{} model.  Units of $\rho$, $\mu_I$ and $T$ are $\mathrm{MeV}$.}
\label{fig:isoRho3D}
\end{figure}

Figures \ref{fig:isoDelta3D} and \ref{fig:isoRho3D} show the solutions $\Delta(T, \mu_I)$ and $\rho(T, \mu_I)$ for the \chiM{} model (note the different axis orientations in the two plots). These plots are similar for the PQM model. We clearly see that no pion condensation occurs for low $\mu_I$. 

\begin{figure}[htb]
  \includegraphics[width=0.4\textwidth,
    , clip=true]{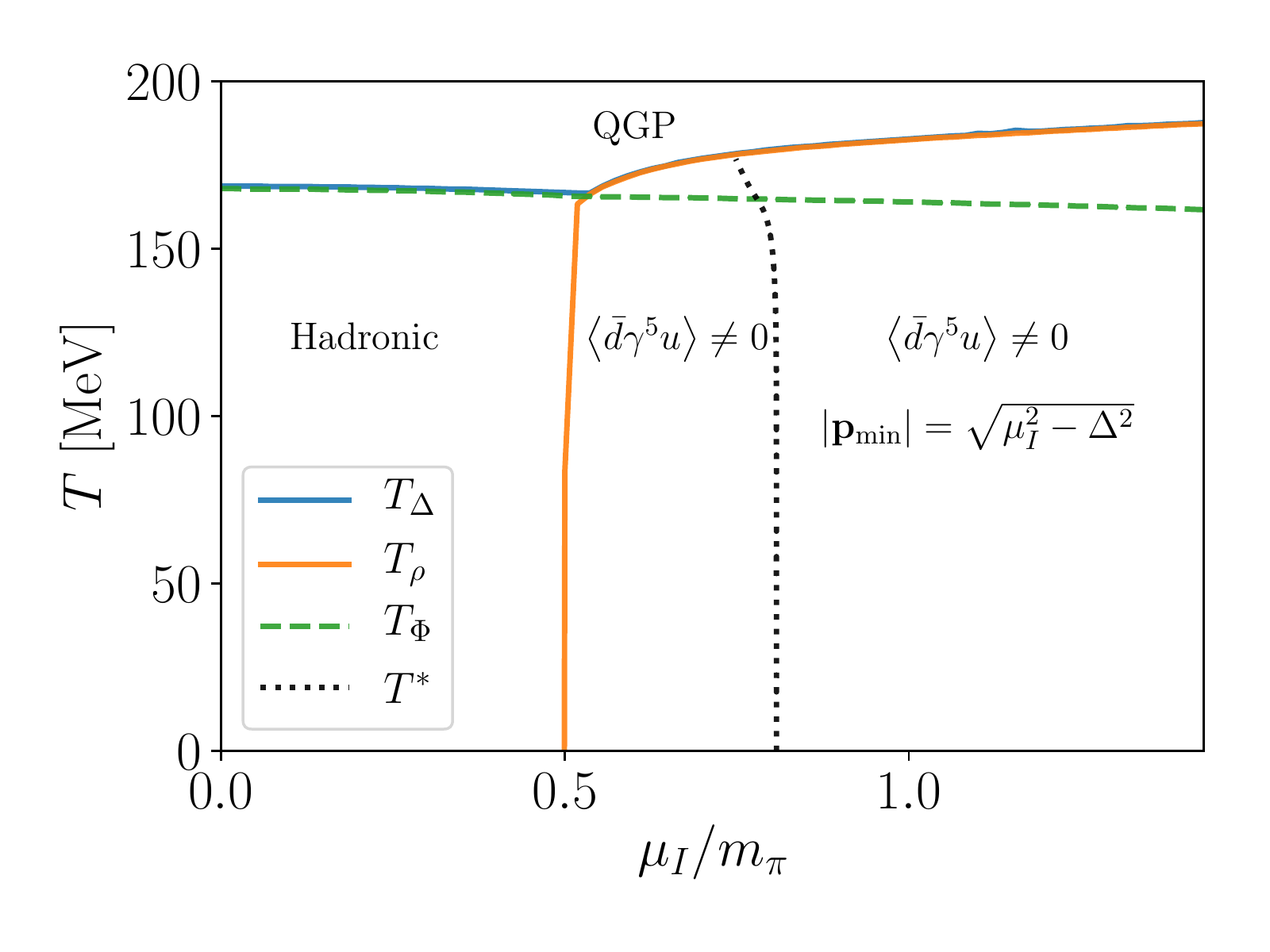}
\caption{Phase diagram in the $\mu_I$--$T$ plane for $\mu_B=0$
for the PQM model. See main text for details.}
\label{phase0}
\end{figure}

\begin{figure}[htb]
\includegraphics[width=0.4\textwidth]{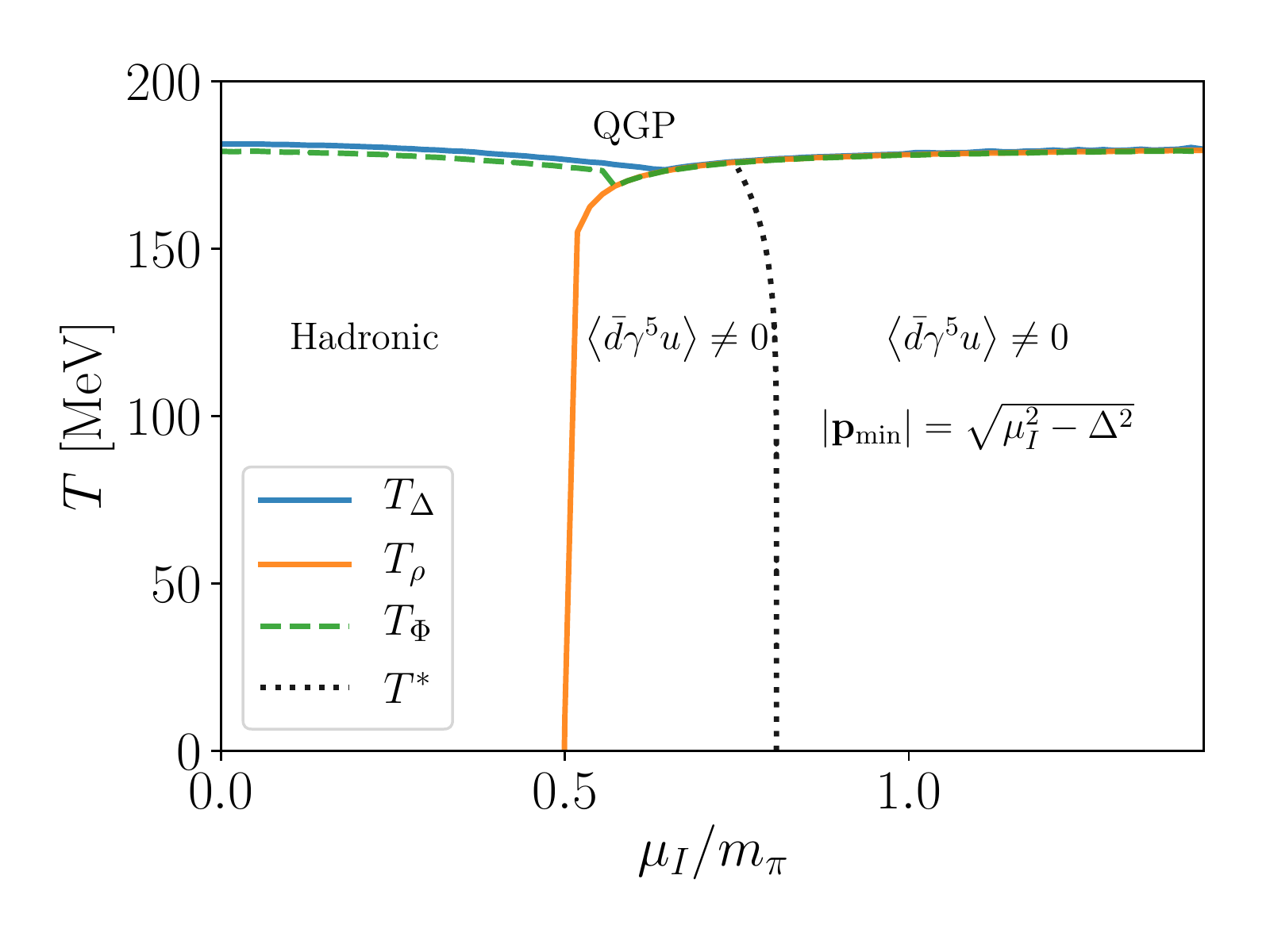}
\caption{Phase diagram in the $\mu_I$--$T$ plane for $\mu_B=0$
for the $\chi M$ model. See main text for details.}
\label{phase1}
\end{figure}

In Figs. \ref{phase0} and \ref{phase1} we show the phase diagram
in the $\mu_I$-$T$ plane obtained in the PQM model and in the $\chi M$ model,
respectively. As mentioned above, the onset of charged pion Bose-Einstein condensation (BEC)
at $T=0$ is at $\mu_I^c={1\over2}m_{\pi}$. The orange line
shows the critical line for BEC, which is fairly steep before it levels off.
The corresponding transition is second order everywhere with
mean-field critical exponents  for the $O(2)$ model.
The transition line for the chiral transition (blue line)
merges with the BEC line at $(\mu_I,T)\approx(\SI{75}{\mega\electronvolt}, \SI{166}{\mega\electronvolt})$ for the PQM model and  $(\mu_I,T)\approx(\SI{90}{\mega\electronvolt}, \SI{173}{\mega\electronvolt})$ for the \chiM{} model. In the $\chi M$ model, the transition line for deconfinement is coinciding with  that of the
chiral transition for nearly all $T$, and consequently, it too merges with the
BEC transition line. Finally, we have drawn a black dashed line within the $O(2)$-symmetry
broken phase. This line is defined by $\mu_I=\Delta$ and starts at $(\mu_I=\SI{113}{\mega\electronvolt}, T=0)$ for both models.
At this value of $\mu_I$, a Fermi surface appears \cite{He_BECBCS}. 
Furthermore, to the right of this line we have $\mu_I>\Delta$, and the energies 
for the $u$ and $\bar{d}$ quarks
\eqref{eu} and \eqref{ed} are no longer minimized by $|\pp|=0$, but rather by $|\pp| = \sqrt{\mu_I^2 - \Delta^2}$. This can been seen as a signal of a transition to a BCS state \cite{sun,tomas, He_BECBCS}. However, we do not have a thermodynamic phase transition, since the same $O(2)$ symmetry is broken on both sides of the dashed line.

In Refs.~\cite{{gergy1,gergy2,gergy3}} the phase diagram in the $\mu_I$--$T$ plane is mapped out
using lattice methods for 2+1 flavors. The phase diagram in Fig.~\ref{phase0} is in especially good 
agreement with that obtained from the lattice: The chiral and deconfinement transition lines
coincide for small values of the chemical potential and meet the BEC transition line
at $(\mu_I^{\rm meet},T^{\rm meet})$. For chemical potentials larger than $\mu_I^{\rm meet}$,
the BEC and chiral lines coincide. Finally, the deconfinement line penetrates smoothly into
the BEC phase. The authors of Refs.~\cite{{gergy1,gergy2,gergy3}} identify this line inside
the $O(2)$-broken phase as the BEC-BCS transition line. Again, the same $O(2)$
symmetry is broken on either side of that line.

We finally note that the phase diagram in Fig.~\ref{phase0} (and also Fig.~\ref{phase1} with the exception of the deconfinement line) seems to agree well with the qualitative phase diagram sketched in Ref.~\cite{quarksonic} based on a large--$N_c$ analysis. They identify the region below the deconfinement line at large $\mu_I$ as a ``quarksonic'' phase where the pressure goes as $\mathcal{O}(N_c^1)$. This phase is argued to be separated by a crossover from the BEC phase where the pressure scales as $\mathcal{O}(N_c^0)$. For a more complete study of this phase with the \chiM{} and PQM models it would be useful to include mesonic fluctuations.

\section{Summary}
In this paper, we extended the chiral matrix model of Pisarski and 
Skokov~\cite{PisarskiSkokov} to finite
baryon and isospin chemical potential.
For temperatures up to approximately up to $2T_c$ and baryon chemical potentials up to $\mu_B=\SI{400}{\mega\electronvolt}$,
this model and the PQM model show reasonable agreement 
with lattice results for a number of thermodynamic functions. 
However, the Polyakov loop rises faster with temperature than on the lattice.
A significant difference between the models was found in the deconfinement phase
diagram. In the $\chi M$ model the deconfinement transition also goes from a 
crossover to a first order transition, with the critical point located at the 
same point as the critical point for the chiral transition. 
This is not the case in the PQM model, where the
deconfinement transition is a crossover for all $\mu$. 
Furthermore, the chiral matrix model predicts deconfinement in the low-$T$, 
large-$\mu$ regime, while the PQM model predicts a 
quarkyonic phase. Thus, the two models predict 
different phases of matter in the low-temperature, high-density regime,
which is the most significant difference between the two models.

Regarding pion condensation at finite temperature, the
two models predict essentially the same phase diagram; the only difference
is that the deconfinement transition merges with the other lines
at large chemical potentials in the \chiM{} model, while in the PQM model the deconfinement line penetrates into the BEC/BCS phase. The phase diagram is in overall good agreement
with the lattice results of \cite{gergy1,gergy2,gergy3}.

\section{Acknowledgments}
The authors would like to thank R. D. Pisarski for 
helpful discussions and the CP-PACS collaboration for providing their lattice data. 
We thank U. Reinosa, J. Maelger and J. Serreau for clarifying details about the imaginary-$r$ saddle point approach.  

\appendix
\section{Thermodynamic potential at one loop in the large-$N_c$ limit}
The tree-level mesonic potential is after symmetry breaking
\bqa
\mathcal{U}_{\rm tree}&=&
{1\over2}{m^2\over g^2}\Delta^2+{\lambda\over24g^4}\Delta^4
-{h\over g}\Delta\;,
\eqa
where we have introduced $\Delta=gv$.
\begin{widetext}
The one-loop contribution to the thermodynamic potential is
\bqa
-    {\log Z\over V\beta}&=&-
    4N_c\int{d^3p\over(2\pi)^3}\omega_p
    -4\int{d^3p\over(2\pi)^3}\left\{{\rm tr_c}\log\left[1+Le^{-\beta(\omega_p-\mu)}\right]
      +{\rm tr_c}\log\left[1+\bar{L}e^{-\beta(\omega_p+\mu)}
      \right]\right\}\;,
      \eqa
      where $\omega_p=\sqrt{\pp^2+\Delta^2}$.
      The first integral on the right-hand side is the quark one-loop contribution to the vacuum potential
      ${\cal U}_{\rm vac}$
      and is divergent for large momenta.
      Using Eq. (\ref{log3d}), we can write
      \end{widetext}
\bqa
    {\cal U}_{q,\rm vac}&=&  {2N_c\Delta^4\over(4\pi)^2}\left[
{1\over\epsilon}+{3\over2} + \ln \frac{\Lambda^2}{\Delta^2} + {\cal O}(\epsilon)
    \right]\;.
    \eqa
    The divergence is eliminated by renormalizing the parameters in the
    Lagrangian. This amounts to 
    making the substitutions
    $m^2\rightarrow Z_{m^2}m^2$, $g^2\rightarrow Z_{g^2}g^2$,
    $\lambda\rightarrow Z_{\lambda}\lambda$, and $h\rightarrow Z_hh$
in the tree-level mesonic potential,
    where~\footnote{In Appendix B, we show that $\Delta=gv$ is not renormalized.}
    \bqa
Z_{m^2}=&\left[1+{4g^2N_c\over(4\pi)^2\epsilon}\right]\;,
\hspace{0.5cm}Z_{g^2}=\left[1+{4g^2N_c\over(4\pi)^2\epsilon}\right]\;,
\\
Z_{\lambda}=&
\left[1+{8g^2N_c\over(4\pi)^2\epsilon}\left(1-{6g^2\over\lambda}\right)\right]\;,
Z_{h}=\left[1+{2g^2N_c\over(4\pi)^2\epsilon}\right]\;.
\eqa
After renormalization, we find the one-loop potential
\bqa\nonumber
\mathcal{U}_{\rm vac}&=&
{1\over2}m_{\ms}^2{\Delta^2\over g_{\ms}^2}
+{\lambda_{\ms}\over24}{\Delta^4\over g_{\ms}^4}-h_{\ms}{\Delta\over g_{\ms}}
\\
&&+{2N_c \Delta^4 \over(4\pi)^2}\left[\log{\Lambda^2\over\Delta^2}+{3\over2}\right]
\;,
\label{potms}
\eqa
where the subscript is a reminder that the renormalized parameters are
in the $\overline{\rm MS}$ scheme.
In Appendix \ref{para}, we show how one can relate the running parameters in the
$\overline{\rm MS}$ scheme to the parameters in the OS scheme and hence
the physical masses and the pion-decay constant.
Substituting the parameters (\ref{runmass})--(\ref{runh})
into Eq. (\ref{potms}). we obtain
Eq.~(\ref{eq:renormGrandPot}).

\section{Parameter fixing}
\label{para}
In this Appendix, we discuss the parameter fixing in the quark-meson model
using the on-shell scheme.
This was first done in Ref.~\cite{crew}.
At tree level, the parameters of the Lagrangian can be
expressed in terms of the phsyical masses and the pion decay constant as
\bqa 
\label{relation1}
m^2&=&-{1\over2}\left(m_{\sigma}^2-3m_{\pi}^2\right)\;,\\
\label{relation2}
\lambda&=&{3}{\left(m_{\sigma}^2-m_{\pi}^2\right)\over f_{\pi}^2}\;,\\
g&=&{m_q\over f_{\pi}}\;,\\
\label{relation3}
h&=&m_{\pi}^2f_{\pi}\;.
\label{relation4}
\eqa
Beyond tree level, these parameters become running parameters in the
$\overline{\rm MS}$ scheme and the relations
(\ref{relation1})--(\ref{relation4}) no longer hold.
The counterterms in this scheme are chosen  such that they exactly cancel
the ultraviolet divergences coming from the loops. 
In the on-shell scheme, the counterterms are chosen
such that they exactly cancel the loop corrections that appear in the
calculations and the parameters therefore still satisfy the above
tree-level relations and are not running. 
Using that the bare parameters in the two renormalization schemes are the
same, we can relate the corresponding renormalized parameters.

The first renormalization condition we impose is that
$\expect{\sigma}=0$,
i.e. that the loop correction to the one-point function
vanishes and that the minimum of the renormalized effective potential
coincides with that of the classical mesonic potential. 
The classical one-point function is denoted by
$\Gamma^{(1)}=it=i(h-m_{\pi}^2v)$ and the classical minimum 
is then given by the equation of motion $t=0$.
Let $\delta\Gamma^{(1)}$ be the one-loop large-$N_c$ correction to the
one-point function. The renormalization condition $\expect{\sigma}=0$
is then
\bqa
\delta\Gamma^{(1)}+i\delta t&=&0\;.
\label{1p}
\eqa

The first on-shell renormalization condition on the two-point function is that
the counterterms exactly cancel the loop corrections that have not
been eliminated by the renormalization condition 
$\expect{\sigma}=0$. 
This gives the mass counterterms
\bqa\nonumber
\delta m_{\sigma}^2&=&i\Sigma_{\sigma}(m_{\sigma}^2)
=8ig^2N_c\bigg[A(m_q^2)
\\ &&
-{1\over2}(m_{\sigma}^2-4m_q^2)B(m_{\sigma}^2)
  \bigg]\;,\\
\delta m_{\pi}^2&=&i\Sigma_{\pi}(m_{\pi}^2)
=8ig^2N_c\left[A(m_q^2)-{1\over2}m_{\pi}^2
  \right]\;,
\eqa
where the four-dimensional integrals $A(m^2)$ and $B(m^2)$
have been defined in Eqs. (\ref{adef}) and (\ref{bdef}).
In the on-shell scheme one also takes as a renormalization
condition that the residue of the propagator at the pole mass equals unity.
This implies
\bqa
    {d\Sigma_{\sigma,\pi}(P^2)\over dP^2}\Big|_{P^2=m^2_{\sigma,\pi}}
+\delta Z_{\sigma,\pi}
    &=&0\;,
\eqa
where $Z_{\sigma,\pi}$ is the wavefunction renormalization counterterm.
One finds
\bqa
\delta Z_{\sigma}&=&
4ig^2N_c\left[B(m_{\sigma}^2)+(m_{\sigma}^2-4m_q^2)B^{\prime}(m_{\sigma}^2)\right]\;,
  \\
  \delta Z_{\pi}&=&
  4ig^2N_c\left[B(m_{\pi}^2)+m_{\pi}^2B^{\prime}(m_{\pi}^2\right]\;.
  \eqa
Let us now return to the renormalization condition (\ref{1p}), which
reads
\bqa
0&=&-8g^2N_cvA(m_q^2)+i\delta t\;.
\eqa
The relation $t=(h-m_{\pi}^2)v$ implies upon variation
a relation among the counterterms,
\bqa
\delta t&=&\delta h_{\os}-\delta m_{\pi}^2v-m_{\pi}^2\delta v_{\os}\;.
\label{dd}
\eqa
In order to find $\delta h_{\os}$, we need to compute $\delta v_{\os}^2$.
The one-loop correction to the quark-pion vertex is of order $N_c^0$
and so is the one-loop correction to the quark field, implying $Z_{\psi}=1$.
Consequently, $\sqrt{Z_{\pi}}\sqrt{Z_{g^2}g^2}=1$, or
${\delta g^2\over g^2}+\delta Z_{\pi}=0$.
A similar argument now applies to $m_q=gv$; since the quark mass correction 
at one-loop is of order $N_c^0$, we find $\delta g v+g\delta v=0$.
Combining these relations, we can write Eq. (\ref{dd}) as
\bqa\nonumber
\delta h_{\os}&=&\delta t+v\delta m_{\pi}^2+{1\over2}vm_{\pi}^2Z_{\pi}^{\os}
\\
&=&-2ig^2N_cm_{\pi}^2v\left[B(m_{\pi}^2)-B^{\prime}(m_{\pi}^2)
  \right]\;.
\eqa
We finally use Eqs. (\ref{relation1})--(\ref{relation2})
to find
relations among the corresponding counterterms
\bqa
\delta m^2_{\os}&=&-{1\over2}\left(\delta m_{\sigma}^2-3\delta m_{\pi}^2\right)
\;,\\
\delta\lambda_{\os}              &=&
3{\left(\delta m_{\sigma}^2-\delta m_{\pi}^2\right)\over v^2}
-\lambda
\delta Z_{\pi}^{\os}\;.
\eqa
This yields
\bqa\nonumber
\delta m_{\os}^2&=&-8ig^2N_c\left[
  A(m_q^2)+{1\over4}\left(m_{\sigma}^2-4m_q^2\right)B(m_{\sigma}^2)
\right.\\ &&\left.
  -{3\over4}B(m_{\pi}^2)
  \right]\;,\\ \nonumber
\delta\lambda_{\os}&=&
-{12ig^2N_c\over v^2}
\left[\left(m_{\sigma}^2-4m_q^2\right)B(m_{\sigma}^2)-B(m_{\pi}^2)
  \right]
\\
&&-4i\lambda g^2N_c\left[B(m_{\pi})+m_{\pi}^2B(m_{\pi}^2)\right]\;.
\eqa
The bare parameters in the Lagrangian are independent of the renormalization scheme.
This implies the following relations among the renormalized
parameters in the two schemes
\bqa
m_{\ms}^2&=&m^2+\delta m_{\os}^2-\delta m_{\ms}^2\;,\\
\lambda_{\ms}^2&=&\lambda+\delta \lambda^2-\delta\lambda_{\ms}^2\;,\\
g_{\ms}^2&=&g^2+\delta g_{\os}^2-\delta g_{\ms}^2\;,\\
h_{\ms}^2&=&h+\delta h_{\os}-\delta h_{\ms}\;,\\
v_{\ms}^2&=&v^2+\delta v_{\os}^2-\delta v_{\ms}^2\;,
\eqa
where we have used that $m^2=m_{\os}^2$ etc.
The counterterms in the on-shell scheme consist of a pole in $\epsilon$
plus finite terms. The former is exactly the counterterm in the
$\overline{\rm MS}$ scheme. Moreover, the parameters in the on-shell
scheme are expressed in terms of the physical masses and the pion decay constant.
We can then express the running parameters in the 
$\overline{\rm MS}$ scheme as
\begin{widetext}
\bqa
\label{runmass}
m_{\ms}^2&=&-{1\over2}\left(m_{\sigma}^2-3m_{\pi}^2\right)
+{2N_cm_q^2\over(4\pi)^2f_{\pi}^2}\left[
  \left(m_{\sigma}^2-3m_{\pi}^2\right)\log{\Lambda^2\over m_q^2}
  +4m_q^2+\left(m_{\sigma}^2-4m_q^2\right)F(m_{\sigma}^2)
  -3m_{\pi}^2F(m_{\pi}^2)
  \right]\;,\\ \nonumber
\lambda_{\ms}&=&{3\left(m_{\sigma}^2-m_{\pi}^2\right)\over f_{\pi}}
+{12N_cm_q^2\over(4\pi)^2f_{\pi}^2}\bigg[
2\left(m_{\sigma}^2-m_{\pi}^2-2m_q^2\right)\log{\Lambda^2\over m_q^2}
+\left(m_{\sigma}^2-4m_q^2\right)F(m_{\sigma}^2)
\\ &&
+\left(m_{\sigma}^2-2m_{\pi}^2\right)F(m_{\pi}^2)
+\left(m_{\sigma}^2-m_{\pi}^2\right)m_{\pi}^2+F^{\prime}(m_{\pi}^2)
\bigg]\;,  \\
\label{rung}
g_{\ms}^2&=&{m_q^2\over f_{\pi}^2}
+{4N_cm_q^2\over(4\pi)^2f_{\pi}^2}\left[
  \log{\Lambda^2\over m_q^2}
  +m_{\pi}^2+m_{\pi}^2F^{\prime}(m_{\pi}^2)\right]\;,\\
\label{runh}
h_{\ms}&=&m_{\pi}^2f_{\pi}
+{2N_cm_q^2\over(4\pi)^2f_{\pi}^2}\left[
  \log{\Lambda^2\over m_q^2}
  +m_{\pi}^2-m_{\pi}^2F^{\prime}(m_{\pi}^2)\right]\;,\\
v_{\ms}^2&=&f_{\pi}^2
-{4N_cm_q^2\over(4\pi)^2f_{\pi}^2}\left[
  \log{\Lambda^2\over m_q^2}
  +m_{\pi}^2+m_{\pi}^2F^{\prime}(m_{\pi}^2)\right]\;.
\label{runv}
\eqa
\end{widetext}
We note from Eqs.~(\ref{rung}) and (\ref{runv}) that the product
$\Delta=g v=g_{\ms}v_{\ms}$, i.e. it does not run with the renormalization
scale.

\begin{figure}[htb]
\includegraphics[width=0.45\textwidth]{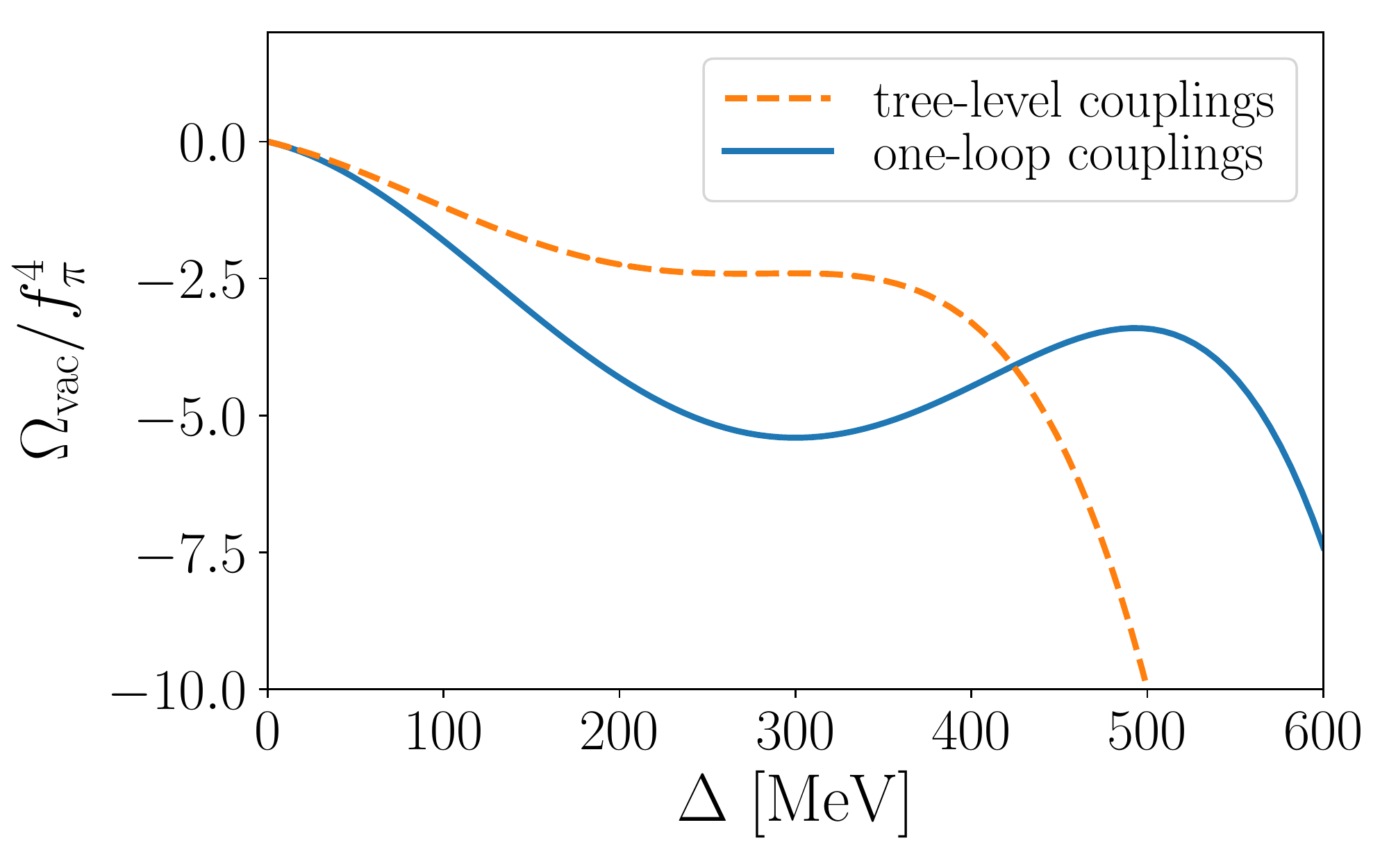}
\caption{Comparison of the effective one-loop potential in the 
  large-$N_c$ limit for $q=r=0$, $T=\mu=0$ with tree-level (dashed orange line)
  and one-loop at large $N_c$ (solid blue line) determination of the couplings.}
	\label{fig:vacPotential}
\end{figure}

Substituting Eqs.~(\ref{runmass})--(\ref{runv}) into the effective potential
(\ref{potms}), we obtain Eq.~(\ref{eq:renormGrandPot}).
We have emphasized the importance of matching the parameters in the one-loop
large-$N_c$ approximation for consistency. For example, the onset of pion condensation
at $T=0$ takes place only at $\mu_I={1\over2}m_{\pi}$ if the parameters
are determined in the same approximation as the effective potential itself.
Moreover, to show the effects of renormalization, we show in Fig.~\ref{fig:vacPotential}
the one-loop effective potential, with couplings determined 
at tree level (dashed orange line) and one loop at large-$N_c$ (solid blue line) with a
sigma mass of \SI{500}{\mega\electronvolt}. Due to the term
$\propto -\Delta^4\log{\Delta^2\over m_q^2}$ in Eq.~(\ref{eq:renormGrandPot}),
the potential will always be unbounded from below for large values of $\Delta$.
However, only in the case where the parameters consistently have
been determined at the same accuracy of the effective potential, is there
a local minimum such that we actually can study the phase transition at finite
$T$ and $\mu$. Using tree-level matching for $m_{\sigma}=500$ MeV,
leads to a vacuum effective potential that cannot be used.

\newpage
\section{Integrals}
In order to renormalize the PQM and $\chi  M$ models, we need
to evaluate some vacuum integrals in four dimensions.
These integrals are divergent in the ultraviolet and we regularize
them using dimensional regularization in $d=4-2\epsilon$ dimension
and the $\overline{\rm MS}$ scheme.
We define
\bqa
\int_Q&=&\left({e^{\gamma_E}\Lambda\over4\pi}\right)^{2\epsilon}
\int{d^{d}Q\over(2\pi)^{d}}\;.
\label{intdef}
\eqa
The integrals needed are
\bqa
A(m^2)&=&i\int_Q{1\over Q^2+m^2}
={im^2\over(4\pi)^2}\left({\Lambda^2\over m^2}\right)^{\epsilon}
\left[{1\over\epsilon}+1
  \right]\;,
\label{adef}
\\ \nonumber
B(P^2)&=&i\int_Q{1\over[Q^2+m^2][(P+Q)^2+m^2]}
\\
 &=&{i\over(4\pi)^2}\left({\Lambda^2\over m^2}\right)^{\epsilon}
\left[{1\over\epsilon}+F(P^2)
  \right]\;,
\label{bdef}
\\
B^{\prime}(P^2)&=&
{1\over(4\pi)^2}F^{\prime}(P^2)\;,
\eqa
where the functions are 
\bqa
\label{fdef}
F(P^2)&=&2-2s\arctan(\mbox{$1\over s$})\;,\\
F^{\prime}(P^2)&=&
{4m_q^2\over P^4s}\arctan(\mbox{$1\over s$})
-{1\over P^2}\;,
\label{fpdef}
\eqa
and $s=\sqrt{{4m_q^2\over P^2}-1}$.

We also need some three-dimensional integrals. In this
case the integrals are defined as in Eq.~(\ref{intdef}) but now with
$d=3-2\epsilon$ instead of $d=4-2\epsilon$. We also use $q$ instead
of $Q$ as integration variable to distinguish the two cases.
The integrals are
\bqa\nonumber
\int_p\sqrt{p^2+m^2}&=&-{m^4\over2(4\pi)^2}
\left({\Lambda^2\over m^2}\right)^{\epsilon}
\left[
{1\over\epsilon}+{3\over2}+{\cal O}(\epsilon)
\right]\;,
\\ &&
\label{log3d} \\
\int_p{1\over(p^2+m^2)^{3\over2}}
&=&
{4\over(4\pi)^2}\left({\Lambda^2\over m^2}\right)^{\epsilon}\left[{1\over\epsilon}+{\cal O}(\epsilon)\right]\;.
\eqa

\section{Propagator and Polyakov loop}
\label{heavyprop}
In this appendix, we show the relation between the
fermion propagator and the Polyakov loop in the nonrelativistic limit, i.e.
for heavy quark masses. We follow Lowell and Weisberger~\cite{nrlimit}
to construct the nonrelativistic limit of the fermion sector in QCD.
We first define the operator $U$ as
\bqa
U&=&\exp\left[-{i\over2m}\gamma^jD_j\right]\;,
\eqa
where the sum over latin indices is only over spatial components.
We also define a new fermion field $\Psi$ via
\bqa
\psi=U\Psi\;.
\eqa
It can be easily shown that the operator $U$ is unitary. 
The quark part of the Lagrangian can then be expanded in powers of
$m^{-1}$ as
\bqa\nonumber
{\cal L}_q&=&\bar{\psi}\left[i\gamma^{\mu}D_{\mu}-m\right]\psi
\\ \nonumber
&=&\Psi^{\dagger}U^{\dagger}\gamma^0\left[i\gamma^0D_0-i\gamma^jD_j-m
\right]U\Psi\\ \nonumber
&=&\Psi^{\dagger}\left[1+{i\over2m}\gamma^jD_j+...\right]
\gamma^0\left[i\gamma^0D_0-i\gamma^jD_j-m\right]
  \\ \nonumber
  &&\times\left[1-{i\over2m}\gamma^jD_j+...\right]\Psi+{\cal O}(m^{-1})
  \\&=&
  \Psi^{\dagger}\left[
-\gamma^0m+iD_0
    \right]\Psi+{\cal O}(m^{-1})\;.
\label{heavylag}
  \eqa
  We next define
  \bqa
  \Psi&=&
  \begin{pmatrix}
	q \\ \tilde{q}^\dagger
  \end{pmatrix},
  \label{eq:psiForm}
  \eqa
  where $q$ and $\tilde{q}^\dagger$ are column $N_c$-plets and
  the upper and lower two-component
  spinors of $\Psi$.
If we use the Dirac representation of the gamma matrices in which
$\gamma^0={\rm diag}(1,1,-1,-1)$, the Lagrangian (\ref{heavylag}) can be
written as 
\begin{equation}
  \lag_{\psi} =  q^\dag(-m + i\partial_t + gA_0^a T^a)q + \tilde{q}(m + i\partial_t
  +
 gA_0^a T^a)\tilde{q}^{\dag}\;.
	\label{eq:NRQCD}
\end{equation}
In the Dirac representation, the upper and lower components of the
Dirac spinors can be interpreted as the particle and antiparticle,
and Eq.~(\ref{eq:NRQCD}) shows that the quark and antiquark degrees of freedom
decouple in this limit.
To get the Lagrangian into the final form, we use
$\tilde{q}T^a \tilde{q}^{\dagger} = \tilde{q}_i T^a_{ij} \tilde{q}^{\dagger}_j = - \tilde{q}_j^{\dagger} T_{ij}^a  \tilde{q}_i =  (\tilde{q}^{\dagger})^T \tilde{T}^a \tilde{q}^T$, where we have defined  $\tilde{T}^a = - (T^a)^T$.
A partial integration yields
$\tilde{q}\partial_t \tilde{q}^{\dag} \simeq -(\partial_t \tilde{q}) \tilde{q}^\dag = (\tilde{q}^\dag)^T \partial_t \tilde{q}^T$, and
redefining $\tilde{q}^\dag$ to be a row object and $\tilde{q}$ a column object,
ie.\ $(\tilde{q}^\dag)^{T} \rightarrow \tilde{q}^\dag$ and $\tilde{q}^T
\rightarrow \tilde{q}$,
the Lagrangian becomes 
\begin{align}
  \lag_{\psi} &=  q^\dag(-m + i\partial_t + gA_0^a T^a)q +
  \tilde{q}^{\dag}(-m + i\partial_t + gA_0^a \tilde{T}^a)\tilde{q} \nonumber \\
  &=  q^\dag D q + \tilde{q}^{\dag} \tilde{D} \tilde{q}\;,
	\label{eq:NRQCD2}
\end{align}
where we have defined the operators
\begin{align}
D = -m + i\partial_t + gA_0^a T^a \label{eq:quarkDiffOp}\;, \\
\tilde{D} = -m + i\partial_t + gA_0^a \tilde{T}^a\;.
\end{align}
Finally, the Hamiltonian density is given by 
\begin{align}\nonumber
  \ham_{q} \equiv& iq^\dag \partial_t q + i\tilde{q}^\dag \partial_t \tilde{q} -
  \lag_{\psi} \\ =& q^{\dag}\left(m-gA_0^a T^a\right)q + \tilde{q}^\dag\left(m - g A_0^a \tilde{T}^a\right)\tilde{q}.
\end{align}
The quark Hamiltonian thus is  
\begin{equation}
	H_q = \int \dd^3 x \mathcal{H}_{q}\;. 
	\label{eq:quarkHam}
\end{equation}

We now want to evaluate the quantity
$\bra{q_a(\xx,0)}\ket{q_a(\xx,-i\beta)}$, which is the
zero-temperature Green's function,
\begin{equation}
[G(\xx, t; \xx, 0)]_{aa} = \bra{q_a(\xx,0)} e^{-iH_q t} \ket{q_a(\xx,0)}, 
\label{eq:Gprop}
\end{equation}
analytically continued to imaginary time $t = -i\tau$ with $\tau = \beta$ for 
a quark state evolving under $H_q$. 
Furthermore, $A_0$ contained in $H_q$ is a classical 
background field.
Since $\lag_q$ is quadratic in the quark fields, the propagator is in practice 
given by a free quantum field theory. The propagator for 
a quadratic Lagrangian $\lag = q^\dag D q$ is the solution to the equation 
\begin{equation}
DG(\xx, t; \xx', 0) = i \delta(\xx-\xx') \delta(t).
\label{eq:propdefeq}
\end{equation}
With $D$ as defined in \eqref{eq:quarkDiffOp}, we find 
\begin{equation}
\left[i\partial_t + gA_0^a(\xx, t) T^a - m\right]G(\xx, t; \xx', 0) = i 
\delta(\xx-\xx') \delta(t)\;.
\end{equation}
When the delta functions are zero, this is just the Schr\"{o}dinger equation, 
which for a time-dependent Hamiltonian $H(t)$ has the well known solution 
\begin{equation}
\timeorder e^{-i\int_0^t \dd t H(t)} =  e^{-imt} \timeorder 
e^{ig \int_0^t \dd t A_0^a(\xx, t) T^a}\;,
\end{equation}
where $\timeorder$ is the time ordering operator. 
With the delta functions included we see by insertion that a solution is 
given by
\begin{equation}
G(\xx, t; \xx', 0) =  \theta(t) \delta(\xx-\xx') e^{-imt} \timeorder e^{ig \int_0^t 
\dd t A_0^a(t) T^a},
\end{equation}
where $\theta(t)$ is the Heaviside step function. This is the retarded 
propagator, which we have chosen since we work in the non-relativistic limit.

In analytically continuing this formula to imaginary times, we will have that that $G(-i\tau, \xx; \xx', 0) = 0$ for imaginary times $\tau < 0$. This is because we should analytically continue to imaginary time before carrying out the path integral implicit in \eqref{eq:Gprop}
to get an analogue of \eqref{eq:propdefeq} in imaginary time. 
We find 
\begin{align}
G(\xx, -i\beta; \xx', 0) &= \delta(\xx-\xx')e^{-\beta m} \timeorder_{\tau} 
e^{ig\int_0^{-i\beta}\dd t A_0^a(\xx, t) T^a}, \nonumber \\
				 &= \delta(\xx-\xx')e^{-\beta m} \timeorder_{\tau} 
e^{g\int_0^{\beta}\dd \tau A_0^a(\xx, -i\tau) T^a},
\end{align}
where we used that $\beta > 0$ and defined the imaginary time ordering operator 
$\timeorder_{\tau}$. 
Defining the Polyakov loop to be 
\bqa
L(\xx) = \timeorder_{\tau} \exp\left[ig \int_0^{\beta} \dd \tau A_4^a(\xx, \tau)
  T^a\right]\;,
\eqa
and introducing the Euclidean gauge field, 
$A_4^a(\xx, \tau) = -iA_0^a(\xx, -i\tau)$, we obtain 
\bqa\nonumber
\bra{q_a(\xx,0)}\ket{q_a(\xx,-i\beta)} 
&=& \left[G(\xx, -i\beta; \xx, 0)\right]_{aa} 
\\
&=& 
V^{-1} e^{-\beta m} \left[L(\xx)\right]_{aa}\;,
	\label{eq:quarkPropagator_appendix}
\eqa
where we have used that $\delta({\bf x}=0)={V}^{-1}$.
Thus the fermion propagator analytically continued to imaginary times is
proportional to the Polyakov loop. The vanishing of the latter
implies the vanishing of the former and is taken as a definition of
confinement.


\bibliography{refs}{}

\bibliographystyle{apsrmp4-1}


\end{document}